\documentclass[preprint2]{aastex1}

\usepackage{natbib}
\bibpunct{(}{)}{;}{a}{}{,}
\usepackage{graphicx}
\usepackage{amsmath}

\shorttitle{Precision pointing of IBEX-Lo observations}
\shortauthors{H{\l}ond, Bzowski, M\"obius et al. }

\begin{document}

\title{Precision pointing of IBEX-Lo observations}

\author{M. H{\l}ond\altaffilmark{1}, M. Bzowski\altaffilmark{1}, E. M{\"o}bius\altaffilmark{2}, H. Kucharek\altaffilmark{2}, D. Heirtzler\altaffilmark{2}, N.A. Schwadron\altaffilmark{2}, M.E.~O\!\'\!\!~Neill\altaffilmark{2}, G. Clark\altaffilmark{2}, G.B. Crew\altaffilmark{3}, S. Fuselier\altaffilmark{4}, D.J. McComas\altaffilmark{5,6}}

\altaffiltext{1}{Space Research Centre of the Polish Academy of Sciences, 18A Bartycka, 00-716 Warsaw, Poland; mhlond@cbk.waw.pl}
\altaffiltext{2}{Space Science Center and Department of Physics, University of New Hampshire, Morse Hall, 8 College Road, Durham, NH 03824, USA; eberhard.moebius@unh.edu}
\altaffiltext{3}{Haystack Observatory, Massachusetts Institute of Technology, Route 40, Westford, MA 01886, USA; gbc@haystack.mit.edu}
\altaffiltext{4}{Lockheed Martin, Space Physics Lab, 3251 Hanover Street, Palo Alto, CA 94304, USA; stephen.a.fuselier@linco.com}
\altaffiltext{5}{Southwest Research Institute, P.O. Drawer 28510, San Antonio, TX 78228, USA; DMcComas@swri.edu}
\altaffiltext{6}{University of Texas at San Antonio, San Antonio, TX 78249, USA; DMcComas@swri.edu}

\begin{abstract}
Post-launch boresight of the IBEX-Lo instrument onboard the Interstellar Boundary Explorer (IBEX) is determined based on IBEX-Lo Star Sensor observations. Accurate information on the boresight of the neutral gas camera is essential for precise determination of interstellar gas flow parameters. Utilizing spin-phase information from the spacecraft attitude control system (ACS), positions of stars observed by the Star Sensor during two years of IBEX measurements were analyzed and compared with positions obtained from a star catalog. No statistically significant differences were observed beyond those expected from the pre-launch uncertainty in the Star Sensor mounting. Based on the star observations and their positions in the spacecraft reference system, pointing of the IBEX satellite spin axis was determined and compared with the pointing obtained from the ACS. Again, no statistically significant deviations were observed. We conclude that no systematic correction for boresight geometry is needed in the analysis of IBEX-Lo observations to determine neutral interstellar gas flow properties. A stack-up of uncertainties in attitude knowledge shows that the instantaneous IBEX-Lo pointing is determined to within $\sim 0.1\degr$~in both spin angle and elevation using either the Star Sensor or the ACS. Further, the Star Sensor can be used to independently determine the spacecraft spin axis. Thus, Star Sensor data can be used reliably to correct the spin phase when the Star Tracker (used by the ACS) is disabled by bright objects in its field-of-view. The Star Sensor can also determine the spin axis during most orbits and thus provides redundancy for the Star Tracker.  
\end{abstract}


\section{Introduction}

One of the key scientific objectives of the Interstellar Boundary Explorer (IBEX) mission \citep{mccomas_etal:09a} is the accurate determination of the interstellar neutral gas flow direction during the key observation period from December through March each year with the IBEX-Lo sensor \citep{mobius_etal:09a} in order to deduce the interstellar (IS) flow vector outside the heliosphere using gravitational deflection of the flow by the Sun. To arrive at results that are at least commensurate in measurement accuracy with the previous determination of the helium IS flow vector by \citet{witte:04} the pointing direction of the IBEX-Lo sensor boresight must be determined with an accuracy better than 0.2\degr~in all directions.

Placing such a requirement on the overall tolerance budget between the spacecraft attitude control and the mounting of the IBEX-Lo sensor, including the entire stack-up of mechanical tolerances and thermal settling of spacecraft components, would have demanded significant resources that were not available. Therefore, a relatively simple Star Sensor was included in the IBEX-Lo sensor package \citep{fuselier_etal:09b}. In this way, the inflow direction of the interstellar neutrals can be directly tied to the positions of stars in the sky as an absolute astronomical reference frame. In order to achieve the required co-alignment solely through careful mechanical assembly, the Star Sensor was mounted on a shared baseplate that also holds the IBEX-Lo collimator, which determines the direction of the incoming neutral atoms.

As a simple Star Sensor concept, we use an optical system with a combination of split-V and pinhole aperture, which is often used for Sun sensors \citep{duchon_vermande:80a, kim_etal:05a, wang_chun:06a}. Because the Star Sensor is co-aligned with IBEX-Lo, it continuously scans an approximately $8{\degr}$-wide strip (as determined by the Star Sensor aperture) along a great circle in the sky that is at $90\degr \pm 7\degr$ from the Sun. All stars within this strip can potentially be used for position determination for a single orbit, during which the IBEX spin axis orientation is kept constant (\citet{mccomas_etal:09a}; see also discussion in the further part of the paper). As described below, such a scheme allows us to determine the positions of stars in this visible strip from the relative timing of the star signals through the two legs of the split-V aperture relative to the start of each spin, which yields information on the spin phase and the elevation (relative to the spin plane of the satellite). This information can then be related to the stars' position in an astronomical reference frame. Relating the pointing of the Star Sensor to astronomical coordinates aids in the analysis of data from the IBEX-Lo sensor and from the IBEX mission as a whole.  

First and foremost, comparing the observations of several stars during a few sample orbits establishes the absolute pointing direction of the Star Sensor and of the IBEX-Lo boresight in the IBEX coordinate system. Therefore, the pointing of the sensor is known in absolute terms for all time periods when the IBEX Attitude Control System (ACS), including its Star Tracker \citep{scherrer_etal:09a}, is operational according to its specifications. During intermittent outages of the ACS due to blinding of the Star Tracker, which occur during periods when either Earth or the Moon get close to the Star Tracker field of view, the spacecraft loses its synchronization with the astronomical coordinates and the spin phase starts to drift slowly. Because the mounting of the Star Sensor and Star Tracker do not change substantially in flight, the Star Sensor signal can be used to ``despin'' the IBEX data accurately so that the start of each spin is exactly re-aligned with its nominal occurrence, 3\degr~before the IBEX-Lo boresight passes the southern ecliptic pole. Finally, for orbits that bring at least two well identifiable stars into the Star Sensor field of view, the orientation of the spin axis can be determined independently of the IBEX ACS. Therefore, the IBEX-Lo Star Sensor provides partial redundancy for the IBEX ACS system.

In this paper, we demonstrate how the orientation of the IBEX-Lo boresight in the IBEX coordinate system was determined, including its uncertainties relative to the Star Sensor and the IBEX satellite system. We show that the IBEX-Lo boresight direction is known during all IS flow observations to a very high accuracy. We also demonstrate that a relatively simple Star Sensor can serve as a stand-alone attitude determination component for simple spacecraft missions.

 In section 2, we provide a functional description of the Star Sensor, followed by the Star Sensor calibration results for the accuracy of the determination of the pointing direction and its brightness sensitivity for a range of stellar magnitudes in Section 3. Section 4 contains a description of the Star Sensor data and of the methods used to analyze them.  Section 5 discusses the accuracy of spin axis determination from ACS data and presents analysis of the Star Sensor observations, starting  from the determination of observed star positions in the sky in relation to their expected positions based on the IBEX spacecraft attitude determination. Then the spin axis orientation is determined independently from the observation of at least two separate stars during individual IBEX orbits and compared with the IBEX attitude determination. The paper ends with 2 Appendices, in which the Star Sensor Simulation Program is presented and deviations of the spin axis derived from Star Sensor observations in a few individual orbits from the ACS-determined spin axis are plotted. 
  
 \section{Functional Description of the IBEX-Lo Star Sensor}
 
\begin{figure}[t]
		\centering
		\plotone{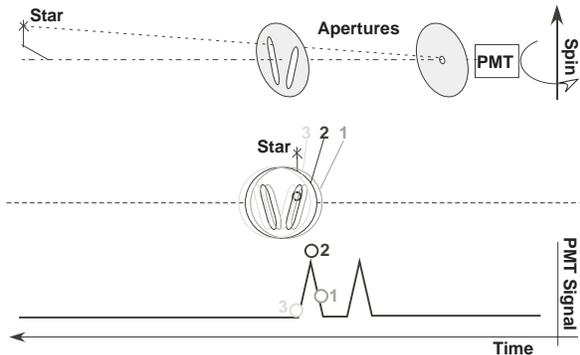}	
  	\caption{Schematic view of the Star Sensor operation. While the satellite spins around the vertical axis, the light from a star reaches the PMT when either one of the legs of the split V is aligned with the pinhole aperture and the star, thus leading to two consecutive near triangular pulses. Top panel: Side view with a star aligned through the right leg.
Center panel: View seen from the PMT through pinhole and split-V aperture for three positions of the split V aperture (1, 2, 3) during a spin. Bottom panel: Light curve elapsing from right to left with positions 1, 2, and 3 from the center panel representations marked.}
 		\label{figSSOpsSchem}
		\end{figure}

 The Star Sensor is to provide true pointing accuracy for the neutral atom observations relative to known star positions in the sky. This is achieved through the use of the timing of the arrival of starlight through a succession of a split-V and pinhole aperture. The general scheme for the operation of the Star Sensor is shown in Fig. \ref{figSSOpsSchem}. Along with IBEX-Lo, the Star Sensor points radially at 90\degr~relative to the spin axis. As the satellite spins, the light from a certain star that is within the narrow angular band swept out by the Star Sensor field of view (FoV) arrive at the detector only when the star, one of the two legs of the split-V, and the pinhole aperture with a photomultiplier (PMT) behind it are aligned. During each of these alignment events with a succession of partial and full overlap of the two apertures, as indicated in the view from the PMT through the pinhole aperture in the center panel of Fig. \ref{figSSOpsSchem} for 3 positions, the illumination of the detector behind the pinhole aperture increases and decreases sharply. The resulting shape of the light curve is close to a triangle, as indicated in the bottom panel of Fig. \ref{figSSOpsSchem} with the 3 positions from above inserted.
		
The mean timing of the two consecutive signals relative to the start of each spacecraft spin provides azimuth information in the spacecraft coordinate system, and the timing difference between the two successive signals in the short sequence of a double-pulse, created by star position in the split-V aperture, provides the elevation. In this way consecutive sky circles will overlap due to the re-orientation of the spin axis towards the Sun at perigee of each orbit. During the first 2.5 years of operation, the orbital period of IBEX was kept at 7--8 days, which requires a re-orientation maneuver of typically 6.9\degr--7.9\degr~in each orbit. The Star Sensor is mounted in the shadow of the IBEX-Lo collimator to shield the sensor from sunlight.	
		
		\begin{figure}[t]
		\centering
		
		\plotone{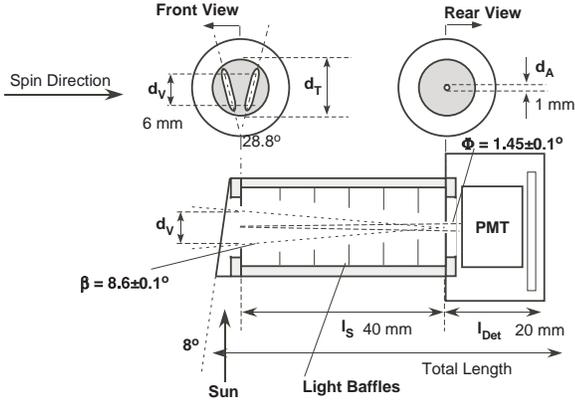}		
  	\caption{Star Sensor scheme, aperture configurations, and relevant linear dimensions.}
 		\label{figSSSchem}
		\end{figure} 
As can be seen from Fig. \ref{figSSOpsSchem} the time difference between the occurrences of the two maxima depends on the actual elevation of the star relative to the Star Sensor boresight. The arithmetic mean of the maxima occurrences determines the crossing of the boresight of the Star Sensor in azimuth or spin angle. The light curve for each star consists of two triangular shaped peaks with a full-width at half maximum (FWHM) that is equivalent to the angular width of the aperture system. Fig. \ref{figSSSchem} shows the aperture layouts for the Star Sensor that has been tailored to its requirements. With an aperture diameter $\mathrm{dA} = 1 \, \mathrm{mm}$, a respective width of 1 mm for the two legs of the split V, its height of 6 mm, and an optical tube length $l_{\mathrm S} = 40$~mm between the two apertures. The FWHM of each of the two signals is $1.45\degr$ and the FoV in elevation from $-5.0\degr$ to $+3.50\degr$. The open diameter to house the aperture system is 11~mm. The pinhole aperture diameter of 1~mm translates into an aperture area of 0.0079~cm$^2$, which transmits $1.23 \times 10^{-15}$~W into the detector for a Sun-like star of magnitude 3. The detector system consists of a compact photomultiplier (PMT) and an amplifier chain provides for light sensitivity down to about magnitude 4. The amplifier chain has a time constant of 12~ms, which shifts the Star Sensor signal by $0.3\degr$ in spin phase, given a nominal spin rate of 4.2~rpm.

The IBEX spin period is typically $14.3\pm 0.2$~s, which is equivalent to $\sim 4.2$~rpm. The PMT signal is accumulated into an array of 720 angular bins with an integration time of $\sim 20$~ms for $0.5\degr$ increments.  While the array has 720 bins, their width is not exactly $0.5\degr$, but generally a little wider, with the last bin of each spin at reduced width to complete $360\degr$. The reason is that the bin width is determined by an integer number $k$ of basic time ticks with a granularity of 69.4 $\mu$s, where $k$ is determined by the actual IBEX spin period as obtained from the IBEX ACS. $k$ is always rounded up to the next integer that makes the bin width just $>0.5\degr$ so that always $\leq 720$~bins are filled with data for one spin. The Star Sensor signal is then accumulated over 64 spins. The absolute reference for this array of 720 bins is the spacecraft spin pulse, whose timing is accurate to 10 $\mu$s (and thus negligible for required angle uncertainties $\leq 0.1\degr$) and which defines the beginning edge of spin bin 0 with high angular accuracy. The spin pulse, which is intended to connect the spacecraft spin phase with an astronomical reference system, is issued when the IBEX-Hi boresight passes $-3\degr$ from the northern ecliptic pole (NEP). Conversely, the IBEX-Lo boresight and thus the Star Sensor boresight passes $-3\degr$ from the southern ecliptic pole at this time.

The Star Sensor data are transmitted as a 720-bin array of 8 bit values every 64 spins, or approximately every 15 minutes with the IBEX-Lo housekeeping data. For the first six months of operation, these data represented actual accumulation over 64 spins. Upon the realization that the IBEX ACS would lose its reference each time when a bright object (Earth or Moon) enters the FoV of the IBEX Star Tracker and the orientation of the spin pulse would drift slowly, a change in the Star Sensor data accumulation was implemented. Because the Star Sensor signals are generally so intense that PMT noise is almost negligible, no accumulation of the signal over consecutive spins is necessary. Therefore, after the first six months of operation (upon completion of the first set of IBEX ENA maps \citep{mccomas_etal:09c}), the data are taken as 1-spin samples for every 64-spin accumulation period.
		
	\section{Calibration of the IBEX-Lo Star Sensor}
Prior to integration into the IBEX-Lo sensor, the Star Sensor was fully tested and calibrated. Two functionally identical models were built and calibrated, the flight model and the flight spare model. Both models showed almost identical characteristics, and, in particular, their pointing, as is relevant to this investigation, turned out to be identical and repeatable within the established pointing accuracy when compared for identical settings during calibrations. Here, we report the calibration results for the flight model (FM). Relevant to this investigation, the FM was calibrated for its response to light equivalent to set star magnitudes, for its angular response and boresight pointing in spin angle and elevation.	

	\subsection{Calibration setup for absolute boresight reference}
The Star Sensor calibrations were performed in a low light calibration facility of the magnetospheric and ionospheric research laboratory of the University of New Hampshire, which contains an Integrating Sphere light source (by Sphere Optics) that is calibrated to NIST standards within a clean dark room. The optics is set up on an optical bench that is isolated from the floor. In order to use the Integrating Sphere, which produces homogenous light over a 10~cm diameter opening, a cover plate with a 4~mm diameter pinhole centered on the opening was added to simulate individual stars. 

The boresight of the Star Sensor was designed and fabricated as perpendicular to the baseplate of the IBEX-Lo sensor, which also defines the boresight of the IBEX-Lo collimator. Through careful choices of machining techniques the mounting of the Star Sensor within IBEX-Lo was controlled such that the collimator boresight is perpendicular to the baseplate to $\pm0.066\degr$.

\begin{figure}[t]
		\centering
		\plotone{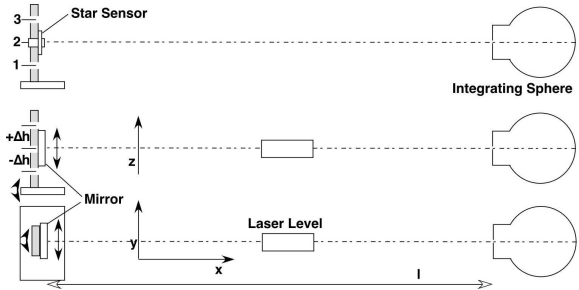}		
  	\caption{Schematic Star Sensor calibration setup. Bottom: Optical setup with a laser level inserted between Integrating Sphere and Star Sensor mount, which contains a precision mirror for alignment in the x-y plane. Center: Setup for alignment in the x-z plane, with the mirror in the center height position $\left(0\degr\right)$. Top: Optical setup with Star Sensor in the center height position.}
 		\label{figSSCalibSetup}
		\end{figure}

In order to calibrate the pointing of the Star Sensor relative to an absolute boresight reference that is perpendicular to its mounting plate a precision jig for the calibration facility was designed, whose perpendicularity could be adjusted and tested on the optical bench with a laser level. A schematic view of the test setup is shown in Fig.~\ref{figSSCalibSetup}. The Star Sensor is mounted on a baseplate in one of three height positions (1, 2, 3) that are separated by $\Delta h = 14$~cm and with position 2 such that the Star Sensor boresight points exactly at the artificial star aperture when correct alignment of the mount is achieved (shown in the top part of Fig.~\ref{figSSCalibSetup}). The separation between the artificial star aperture and the pinhole aperture of the Star Sensor is $l = 248$~cm, thus leading to elevation pointing of $+3.23\pm  0.02\degr$~(1), 0\degr~(2), and $-3.23 \pm 0.02\degr$~(3). The jig is attached to a precision rotation table whose rotation axis represents the z-axis and is centered on the pinhole aperture of the Star Sensor so that the Star Sensor can be rotated relative to the artificial star to simulate the spacecraft spin. For exact alignment, the rotation table is set to 0\degr.

 To achieve alignment a laser level that emits laser light along the same axis to both sides is mounted between the Integrating Sphere and the Star Sensor mount (shown in the center and bottom parts of Fig.~\ref{figSSCalibSetup}). The laser beam adjusts itself to exactly level. The laser level is adjusted in y and z directions so that it is exactly centered on the artificial star aperture. A precision polished metal plate with 4 markings at equal distance from the center of the metal plate is placed into position 2 of the Star Sensor mount. Now the mount is adjusted in the y and z directions so that the laser beam is centered between the 4 center markings. Finally, the mount is adjusted in angle in the x-y and in the x-z plane so that the laser beam is exactly reflected back into the beam exit of the laser level. These alignments are achieved to $\pm 0.5\, \mathrm{mm}$ so that an overall alignment accuracy and thus Star Sensor reference in perpendicularity to the mounting plate of $\leq 0.018\degr$~was achieved.
 
For consistency we used a mirror that was placed at the position of the polished plate which reflects the laser beam back on itself. Hence we double the travel distance of the laser beam, and thus, we doubled the accuracy.
 
 	\subsection{Calibration of angular response and boresight pointing}
\begin{figure}[t]
		\centering
		\plotone{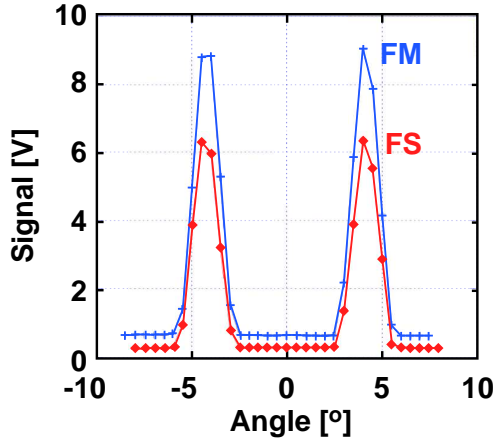} 
  	\caption{Angular scan with the Star Sensor (Flight Model in blue, Flight Spare in red) in elevation position 2 in $0.5\degr$ increments.}
 		\label{figSSangScan1}
		\end{figure}

To calibrate the Star Sensor for its angular response the orientation of the Star Sensor is changed in increments of 0.5\degr~ with the rotation table when the Star Sensor is mounted in either of the three height positions to allow for the simulation of three different elevations. Fig.~\ref{figSSangScan1} shows the result from such a scan for the FM and the FS in comparison with an illumination equivalent to a magnitude 1.5 star. The measurements were obtained with engineering model electronics, which allows an amplifier output range from 0~to 10~V. 0\degr~ on the angle scale refers to exact alignment of the boresight with the line-of-sight to the artificial star. The results in Fig.~\ref{figSSangScan1} demonstrate that the alignment of the Star Sensor was highly reproducible, even between different Star Sensor models. The reproducibility was also achieved before and after vibration testing of the Star Sensor.
 
For exact boresight pointing, the center position in spin angle was verified to be accurate within $\pm0.065\degr$. The separation of the legs of the split V aperture in the center position (2) in the mount, i.e. $0\degr$~elevation, was found as $8.365\pm0.039\degr$. Including the fact that the two legs are inclined at 14.4\degr~relative to the center line, the elevation pointing was found to be accurate to $\pm 0.077\degr$.		

Including the mechanical tolerance stack-up of the collimator boresight of 0.066\degr~and the uncertainty in the knowledge of the alignment of the Star Sensor in the calibration mount of 0.018\degr, the accuracies of the alignment of the Star Sensor and the collimator boresight are known to within 0.094\degr~in spin angle and 0.102\degr~in elevation.

		\subsection{Sensitivity and dynamic range of the Star Sensor}
Both Star Sensor models have been calibrated for a wide range of star magnitudes that include the magnitudes of the visible outer planets (Mars, Jupiter, and Saturn). This calibration provides the necessary information to set the bias voltage of the photomultiplier (PMT) tube adequately in flight and to help identify bright stars in the Star Sensor signals in flight. It should be noted that the star magnitudes are given as magnitudes in the visual spectral band based on the color temperature of the lamp in the Integrating Sphere. To relate the results to actual star magnitudes the spectral curves of individual stars and the spectral curve of the PMT need to be factored in. Because the Star Sensor will be solely used for pointing purposes, no absolute calibration taking into account these spectral sensitivities has been attempted. The result shows the wide dynamic range that the Star Sensor can operate in. In additional tests with a different light source, which was less well calibrated than the Integrating Sphere, it was verified that the Star Sensor is even capable of reliable observations of the half moon (magnitude $\sim-10^{\mathrm m}$) with PMT voltages between 200 and 250~V, which are now used in flight for moon viewing time periods.		
		\section{Data analysis}
		
			\subsection{Description of Star Sensor data}
The ``first light'' observations  from the Star Sensor demonstrated that the instrument is working excellently and very close to expectations. This is illustrated in Fig. \ref{figSSstarIdent} where an observation from a selected orbit is compared to the signal predicted by the Star Sensor Simulation Program (Appendix A). Several effects can obscure Star Sensor data: the signal from the Star Sensor can be saturated by the light of the Moon or Earth, when they are close to the FoV; and the synchronization of the spin pulse can be lost during a portion of the orbit if either the Earth or the Moon blind the spacecraft Star Tracker. The Star Sensor signals are used to correct pointing information when the spin-pulse is lost. 			
		
\begin{figure*}[t]
		\epsscale {2.0}  
		\centering
		\plotone{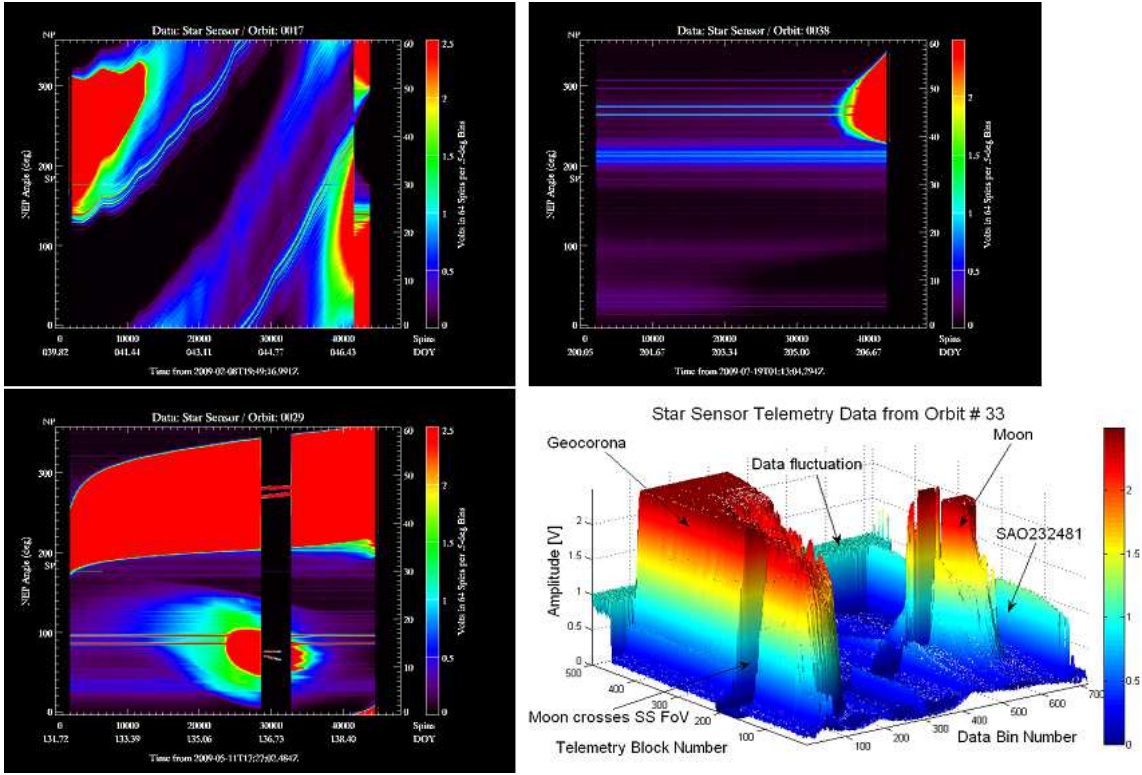}
  	\caption{Examples of data from various orbits, shown to illustrate the complex nature of the Star Sensor signal. The two upper and the lower left panels show a 2D time versus North Ecliptic Pole angle representation of the Star Sensor signal for 3 different orbits. The upper left panel shows an example when the star tracker is in eclipse, which causes the spin-phase to meander from its proper value. The lower-right panel presents a 3D visualization of the Star Sensor telemetry data collected from the entire orbit 33. The intensity of the signal is colour-coded and displayed at the vertical axis. See text for discussion.}
 		\label{figSSTelemData}
		\end{figure*}

Examples of such effects in the Star Sensor data are shown in Fig. \ref{figSSTelemData}. The upper two panels and the lower left panel show the Star Sensor signal as function of spin phase (vertical) and time (horizontal) in a color-coded representation, and the lower-right panel presents an example of the data from orbit 33 in a more detailed 3D view.

In a scenario that is straightforward to analyze, a point source should show up as a pair of horizontal ``rails'', which correspond to the double peak of the star over time. Diffuse background should show up as broad horizontal bands. Any departures from strictly horizontal structures mean that either a time-related phenomenon (like an object moving through the FoV) is observed or a problem with spin synchronization exists. The synchronization problems are illustrated in the upper left panel, where the horizontal structures are transformed into oblique wavy ones -- and the waviness is similar for all elements of the picture. Only at the very end of this orbit, at the far right in the panel, the synchronization was restored. In such a case, the observations of point sources from the Star Sensor can potentially be used to despin the observations provided that at least one clearly visible star is in the FoV.

If the Star Sensor is to be used also to independently determine the spacecraft spin axis orientation, at least two well separable stars are required. The upper right panel shows an example of a good orbit for this purpose: two stars are visible against a relatively weak background in the upper portion of the panel. There is even a possibility to resolve a third star, but this one is seen against a bright background strongly varying with spin phase, which makes using it for the spin axis pointing determination problematic. However, the remaining two stars are sufficient for the task. Only at the end of the orbit, strong reflected stray light from the Earth enters into detector and swamps the two good stars. 

The lower left panel illustrates a case of heavy contamination by the glow of the Earth (the big red structure in the upper portion of the panel) and Moon (the oval structure). There are also two bright stars shining through the background and the Moon glow. At about 2/3 into the orbit interval the Star Sensor was temporarily switched into the Moon Mode (with appropriate reduction of the PMT voltage) because of the expected passage of the Moon through the FoV. This time period is visible as the vertical black strip. And indeed, the Moon signal was registered, as manifested by the two lines visible against the dark background cutting through the oval. The Moon shows a component of motion through the FoV from lower to higher elevation in the S/C reference system. As a consequence, the angular distance between the two peaks is increasing with time. During this particular orbit also the Earth was within FoV simultaneously with the Moon, as evidenced by the strong signal visible during the Moon Mode time period. The Earth signal is saturated even in the Moon Mode, nevertheless the Earth is resolved and visible as a compact object (since it gives two distinct peaks). Because of the orbital motion of IBEX, the Earth is crossing the FoV downwards (i.e. its elevation decreasing, in contrast to the Moon), as can be established from the fact the distance between the two peaks is decreasing with time. It is evident that the Earth and the Moon move also in azimuth, as indicated by the fact that the imaginary center lines between their double peaks are not horizontal, as are the signals from the two stars. Since the synchronization is maintained in this orbit, the Star Sensor provides us with time-resolved positional information on the objects moving through the FoV. Even under such challenging conditions, the signal from the two good stars at the beginning and the end of the orbit could be used for the axis pointing determination. 

This qualitative discussion is offered to justify that given the complex and varying composition of the data we decided not to develop an algorithm for an automatic selection of suitable data intervals for the analysis. Instead, we rely on human judgment supported by simulation.

 		\subsection{Determination of star positions in the spacecraft reference frame}

Determination of orientation of the optical axis of the IBEX-Lo instrument from the Star Sensor observations requires two steps: extraction of positions of point-like objects (stars, planets) from the telemetry data and identification of these objects with stars or planets listed in a stellar catalog or planetary ephemerides for the time of the observation. Typical reference objects most frequently observed by the Star Sensor will evidently be bright stars, but also the naked-eye outer planets and even the Moon can be considered as additional reference sources.

The point-like stars are always observed against a comparably strong diffuse sky background. The most important components of the background are the zodiacal light, diffuse Galactic light and a huge number of faint, unresolved stars \citep{leinert_etal:98a}. Hence the signal registered by the Star Sensor is composed of signals from the objects of interest for the Star Sensor on top of an extended background in the spectral sensitivity range of the detector. 

\begin{figure*}[t]
		\centering
		\plotone{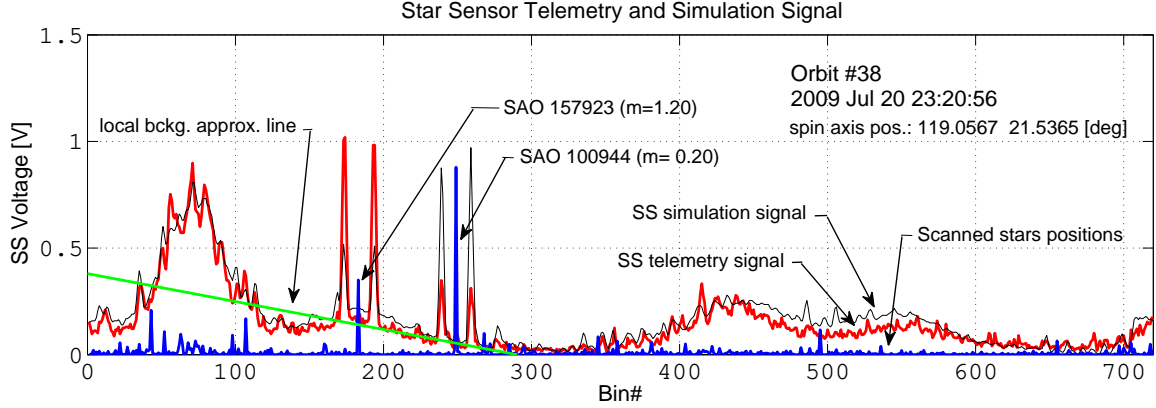}
  	\caption{Illustration of the process of identification of the strongest valid peak pairs in the Star Sensor histogram. The Star Sensor telemetry histogram obtained during orbit 38 on 2009 Jul 20, 23:20:56 is shown as a red line. The simulated signal for the same orbit as obtained from the Star Sensor Simulator program is shown as a black line. The azimuth and brightness of stars brighter than $m = 6$ are shown (blue line). The azimuth of a star is the mean value of the position of two peaks that the Star Sensor returns from a point source in the sky. There are two bright stars with a high signal-to-background ratio in the figure. The star SAO 100944 is observed against a background whose intensity varies within the azimuth range occupied by the double peak. Thus, before processing of the star signal the background must be subtracted, which is done after fitting the background linearly (green line).}
 		\label{figSSstarIdent}
		\end{figure*}

The method used to determine the positions of stars from the IBEX telemetry data is inverse to the method used to simulate the Star Sensor response function, described in Appendix A. The Star Sensor data collected during one entire orbit typically include a few hundred histograms (scan data blocks). To identify stars, we look for double pulses which are considerably stronger than the neighboring sky background level (Fig. \ref{figSSstarIdent}) and do not have other strong stars nearby. Strong neighbor stars in the FoV modify the shape of the pulses of the target star. Because the distortions of each of the two peaks usually differ, the distance between the two peaks is changed, which impacts both the azimuth and elevation obtained from such a star. The selection of the candidate peak pairs is finalized through human involvement because of the varying and complex character of the signal, as discussed in the preceding section.
 
Processing of observations from a given orbit begins with selection of a subset of data blocks free from synchronization issues and saturated reflected light. Then, the histogram portions with the peak pairs are determined. The latter choice is maintained for all suitable data blocks from the given orbit because the stability of the signal is very good for orbits in full spin synchronization. Only peaks stronger than a preselected level of 0.15~V are analyzed.   

 		\begin{figure}[t]
 		\epsscale {1.2}  
		\centering
		\plotone{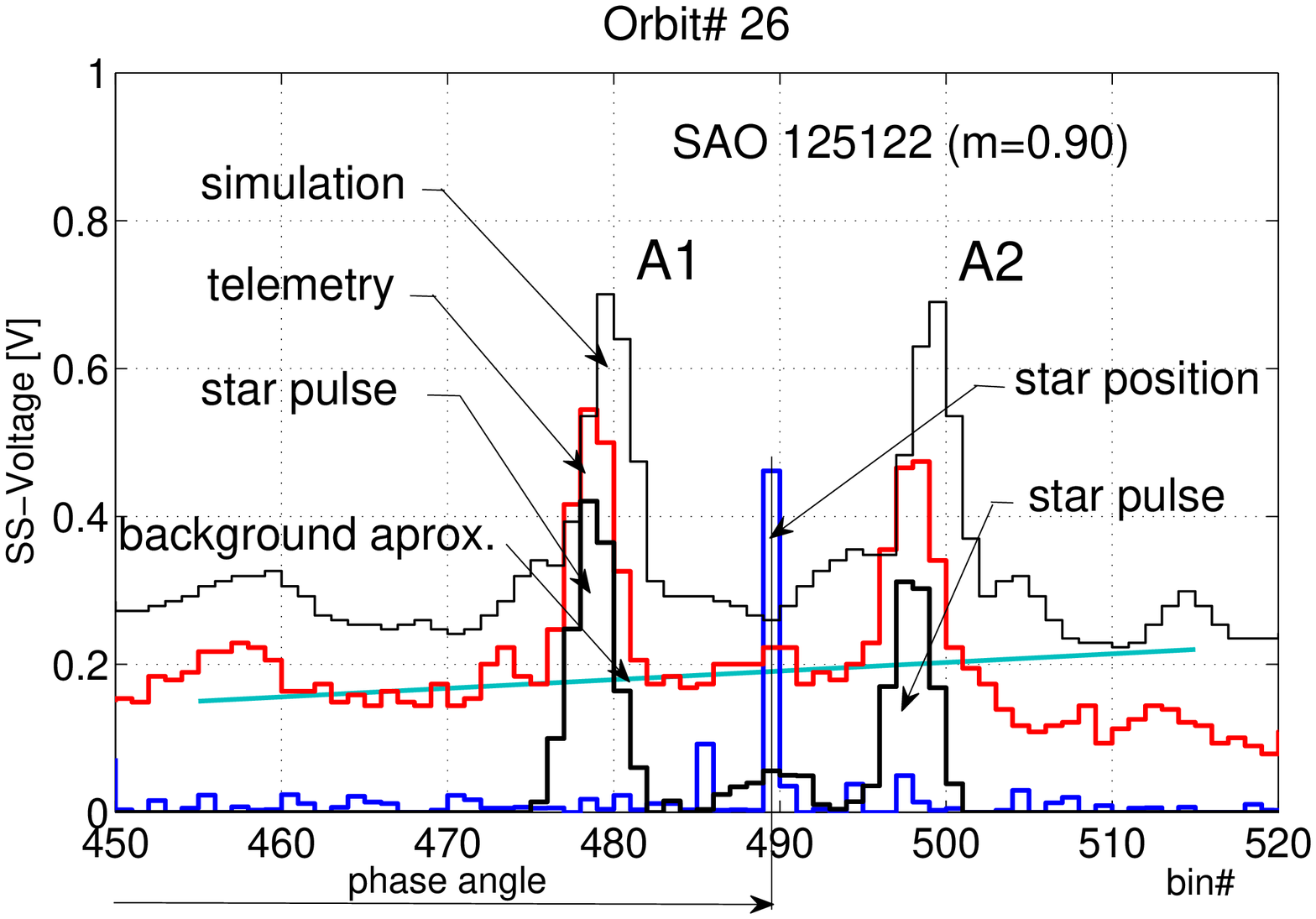}   
  	\caption{Illustration of the method used for determination of the scanned star position in the S/C coordinates system from the Star Sensor histogram data. In the telemetry signal (red line) there are two pulses, A1 and A2, of the Star Sensor output voltage resulting from a star passage through the Star Sensor FoV. They are observed against sky background. With the boundaries of the peak pair in the histogram identified, the peaks are temporarily removed from the histogram, and a linear model is fitted to the remaining background signal. Then the background is subtracted from the original signal, and the resulting peaks are processed to determine their positions and thus the star azimuth. The results are checked against the simulations. The simulation result is superimposed on the data (marked as ``simulation''). The position of the star along with all stars brighter than $m = 6$ is shown in the figure in blue.}
 		\label{figSSPeakPosits}
		\end{figure} 

With the appropriate pairs of peaks selected, the local sky background level (approximated by straight line, Fig. \ref{figSSPeakPosits}) is subtracted from the sensor signal, as illustrated in Figs \ref{figSSstarIdent} and \ref{figSSPeakPosits}. Based on the residual signal, the center position of each peak is determined through the calculation of its center-of-mass. Because the FWHM of the slit width is 1.4\degr, each peak typically extends over seven angular bins. After identifying the maximum bin of a peak, the center-of-mass is obtained from the seven bins centered on this bin, using bin centers as angular coordinates.

We obtain the spin angles $\alpha_1$, $\alpha_2$ for both peaks of a pair. Before further processing of the selected pair we check whether it meets the criteria of the split-V aperture geometry, i.e. 
\begin{equation}
5.73\degr \leq \alpha_2 - \alpha_1 \leq 9.87\degr 
\label{eqSSStarAzim}
\end{equation}
If valid, the azimuth (spin angle) of the identified star $\alpha_{\mathrm{star}}$ in the spacecraft system is derived as
\begin{equation}
\alpha_{\mathrm{star}} = \left(\alpha_1 + \alpha_2\right)/2
\label{eqSSStarAzim1}
\end{equation}
We do not require here that the peaks are equal in height because they still might be affected by unaccounted background and photocathode inhomogeneities.

With the azimuths of the two peaks and the azimuth of the star, the elevation of the star $\delta_{\mathrm star}$ can be calculated from the formula:
\begin{equation}
\delta_{\mathrm{star}} = \arcsin\left[\frac{\tan\left(\left(\alpha_2 - \alpha_1 - \sigma_V \right)/2\right)}{\tan\left(\beta_V \right)} \right]
\label{eqSSStarElev}
\end{equation}
where $\beta_V = 14.4\degr$ is the tilt angle of each slit in the split-V aperture relative to the symmetry line and $\sigma_V = 8.4\degr$ is the angular separation of the slits at 0\degr~elevation. 		

		\subsection{Correction of an instrumental shift in spin angle (azimuth)}

A detailed comparison of the spin angles (azimuths) of stars determined from the telemetry with the spin angles expected from simulation showed that the spin angles of the stars in the S/C frame are shifted relative to the positions expected from simulations by an angle that increases with the spin angle. We identified two reasons for this effect: (1) a constant shift by $\sim 0.3\degr$~is introduced by the Star Sensor amplifier, and (2) a linear ``scaling'' effect in the on-board signal formation process exists. The scaling effect comes up because of a finite resolution of the coding of the duration of the clock ticks used for bin width determination, indicated in Section 2. Here, we develop a correction factor, $f_s$, that is the ratio of the actual bin width divided by the 0.5\degr~bin width. The actual bin width can be solved for using a housekeeping register $k$ that is used to solve for the time interval (in seconds) $\tau_s = (k+192)/14400$ between the interrupts that separate each of the 720 spin-bins. The actual bin-width (in degrees) is then $\Delta \phi_s=360\degr \tau_s/T_{\mathrm{CEU}}$, where $T_{\mathrm{CEU}}$ is the spin period of the S/C stored in the memory of the on-board computer and measured on a spin by spin basis. The correction factor $f_s = \Delta \phi_s/0.5\degr$ or, combining factors:
\begin{equation}
f_s = \left(k + 192\right)\frac{720}{14400}\frac{1}{T_{\mathrm{CEU}}}
\label{equSSPhaseCorr}
\end{equation}
The corrections for the instrumental shift (0.3\degr) and scaling of the spin phases are implemented directly in the program processing the Star Sensor telemetry data.

The method for the star position determination has been extensively tested on the signal obtained from simulations. A series of 50 cycles of the Star Sensor signal simulation/scanned star position finding were performed for a few IBEX orbits (spin axis positions). Since the simulations include statistical fluctuations of the intensity of the signal, the results were statistically distributed, but the mean position of the identified stars were found to be only slightly different from the actual star positions used for the simulation. For example, for IBEX Orbit 22 the error in right ascension was only $\sim 0.02\degr$~and in declination $\sim 0.05\degr$. In this way we made sure that the implemented method could determine the star positions with sufficient accuracy after the input data had been appropriately corrected.
 
  	\begin{figure}[t]
		\centering
		\plotone{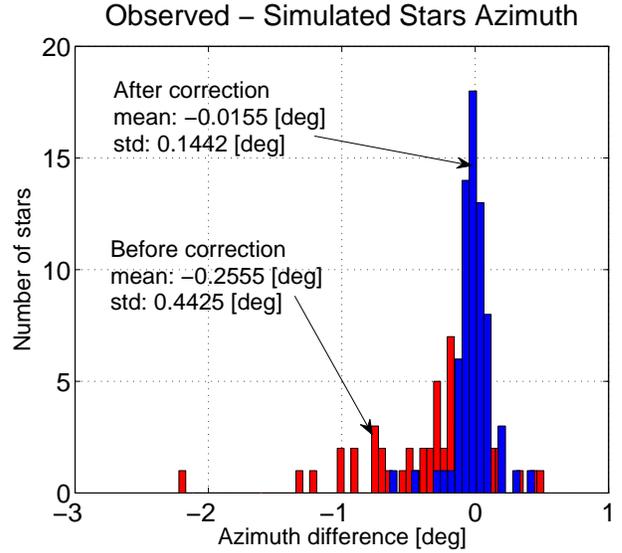}   
  	\caption{Illustration of the effectiveness of phase correction defined in Eq. \ref{equSSPhaseCorr}. Differences between the orbit-averaged observed and simulated azimuth of observed before the correction are shown as the red histogram. The differences after the correction are represented by the blue histogram.}
 		\label{figSSCorrEffect}
		\end{figure} 

An illustration of quality of the correction defined in Eq.~(\ref{equSSPhaseCorr}) is shown in Fig.~\ref{figSSCorrEffect}, where histograms of the differences between the orbit-averaged observed and simulated spin angles (azimuths) of the stars in the S/C frame are shown. The red histogram of pre-correction differences is not symmetrical and features an extended wing towards the negative differences. The differences between the simulated and post-correction azimuths (blue in the figure) are much more symmetric and the histogram is narrower. This histogram is shown in a finer scale in the right-hand panel of Fig. \ref{figSSMeanDev}.		

		\subsection{Identification of found objects with catalog stars}
Once the positions of the stars in the spacecraft reference system are determined, they have to be identified with reference stars from the star catalogue. The identification is based on a comparison of the telemetry data with simulations results (Fig. \ref{figSSstarIdent}). The simulation of the signal expected from the telemetry is calculated for a given UTC time and spin axis position obtained from the Star Tracker signal processed by the ISOC \citep{schwadron_etal:09a}. Based on the simulation results, a short list of candidate stars is returned whose positions are then compared with the observed star positions. Details of the simulation program are given in Appendix A. Even though a software module to identify the stars from the observations was developed and successfully tested, the manual identification turned out to be more practical in situations when the signal is affected by one of the unexpected complexities discussed in the preceding sections. 		

		\subsection{Determination of the spin axis orientation}
If at least two stars for one orbit can be identified with star catalog entries, it is possible to calculate the spin axis orientation. Determining the axis pointing requires finding a matrix $\tens  {R}^{bi}$ that transforms the orientation of the spacecraft reference system, in which we have obtained the star positions, into the target inertial celestial reference system. In our case, the target reference system is the equatorial system and the method adopted is the TRIAD algorithm \citep{shuster_oh:81a}. There is relatively little robustness in the process of determining the S/C attitude using observations of just 2 stars because there is no redundancy of information. Nonetheless, this method offers a very valuable way to compare the Star Sensor's attitude determination with that of the ACS system. Furthermore, the resulting spin axis pointings can be statistically analyzed to validate the actual pointing of the Star Sensor with that of IBEX-Lo in spacecraft coordinates. Finally, it is demonstrated that the Star Sensor could provide the spacecraft attitude determination if such a need would arise.	

In order to determine the spin axis orientation with the use of the TRIAD algorithm, the positions of the two stars must be determined quite accurately, which is satisfied by the selection criteria for the stars as described above. In practice, such conditions are not always fulfilled and thus there are orbits for which no determination of the spin axis position is possible because of the lack of suitable stars.	

The relation between the measured star position vectors in the S/C system $\vec{v}_{1b}$, $\vec{v}_{2b}$ and their position vectors in the equatorial system $\vec{v}_{1i}$, $\vec{v}_{2i}$ is:
\begin{equation}
\vec{v}_{1b} = \tens{R}^{bi}\vec{v}_{1i}\textrm{ and } \vec{v}_{2b} = \tens{R}^{bi} \vec{v}_{2i}
\label{eqSSTRIAD1}
\end{equation}
which must be solved for the transformation matrix $\tens{R}^{bi}$. The matrix algorithm is based on the construction of two triads $\vec{t}_b$ and $\vec{t}_i$ of ortho-normal unit vectors using the observed and reference star position vectors $\vec{v}_{1i}$, $\vec{v}_{2i}$, $\vec{v}_{1b}$, $\vec{v}_{2b}$ \citep{shuster_oh:81a}: 
\begin{equation}
\vec{t}_b = \left[\vec{t}_{1b}, \vec{t}_{2b}, \vec{t}_{3b}\right] \textrm{ and } \vec{t}_i = \left[\vec{t}_{1i}, \vec{t}_{2i}, \vec{t}_{3i} \right]
\label{eqSSTriadDef1}
\end{equation}
where
\begin{eqnarray}
\vec{t}_{1i} &=& \vec{v}_{1i}\nonumber \\ 
\vec{t}_{2i} &=& \frac{\vec{v}_{1i} \times \vec{v}_{2i}}{\left|\vec{v}_{1i} \times \vec{v}_{2i}\right|} \\ 
\vec{t}_{3i} &=& \vec{t}_{1i} \times \vec{t}_{2i}  \nonumber 
\label{eqSSTriadDef2a}
\end{eqnarray}
\begin{eqnarray}
\vec{t}_{1b} &=& \vec{v}_{1b}\nonumber \\ 
\vec{t}_{2b} &=& \frac{\vec{v}_{1b} \times \vec{v}_{2b}}{\left|\vec{v}_{1b} \times \vec{v}_{2b}\right|} \\ 
\vec{t}_{3b} &=& \vec{t}_{1b} \times \vec{t}_{2b}  \nonumber 
\label{eqSSTriadDef2b}
\end{eqnarray}
\begin{eqnarray}
\left[\vec{t}_{1i}, \vec{t}_{2i}, \vec{t}_{3i}\right] &=&  \tens{R}^{it} \tens{I} = \tens{R}^{it} \nonumber \\
\left[\vec{t}_{1b}, \vec{t}_{2b}, \vec{t}_{3b}\right] &=&  \tens{R}^{bt} \tens{I} = \tens{R}^{bt}
\label{eqSSTriadDef2c}
\end{eqnarray}
Finally:
\begin{eqnarray}
\tens{R}^{bi} &=& \tens{R}^{bt}\left(\tens{R}^{it}\right)^{\mathrm{T}} \nonumber \\
              &=& \left[\vec{t}_{1b}, \vec{t}_{2b}, \vec{t}_{3b}\right] \left[\vec{t}_{1i}, \vec{t}_{2i}, \vec{t}_{3i} \right]^{\mathrm{T}}
\label{eqSSRBIDef}
\end{eqnarray}
represents the transformation from the equatorial system into the S/C system.  The inverse transformation is the transposed matrix:
\begin{equation}
\tens{R}^{ib} = \left(\tens{R}^{bi} \right)^{\mathrm{T}}
\label{eqSSRIBDef2}
\end{equation}

The IBEX spin axis is the z-axis of the S/C reference system. Therefore, the transformation matrix allows the calculation of the spin axis $\vec{v}_{zb} = \left[0, 0, 1 \right]$ in the equatorial system as:
\begin{equation}
\vec{v}_{zi} = \left(\tens{R}^{bi}\right)^{\mathrm{T}}\vec{v}_{zb}
\label{eqSSSpinAxDef}
\end{equation}

The method was implemented numerically and successfully applied to determine the spin axis pointing of IBEX during the first two years of science operations. 

	\section{Results}
	
		\subsection{Accuracy of spin axis pointing from ACS}
Pointing of the IBEX spin axis is determined by the ISOC based on data from the IBEX ACS. Experience from more than 2 years of operations showed that when the observation conditions are favorable, the ACS performs with high precision and according to specifications, enabling a very accurate determination of spin axis pointing. The spin axis is very stable during an entire orbit, with a very small precession amplitude of $\sim 0.02\degr$, i.e. comparable to the typical precision of the ACS system. The high stability of the IBEX spin axis enables us to adopt one solution for the entire duration of each orbit and to perform a reliable statistical comparison with solutions obtained by the Star Sensor, as discussed in the following sections. 
	
		\subsection{Comparison of the predicted and measured positions of stars}
\begin{figure*}[t]
		\epsscale {1.8}  
		\centering
		\plottwo {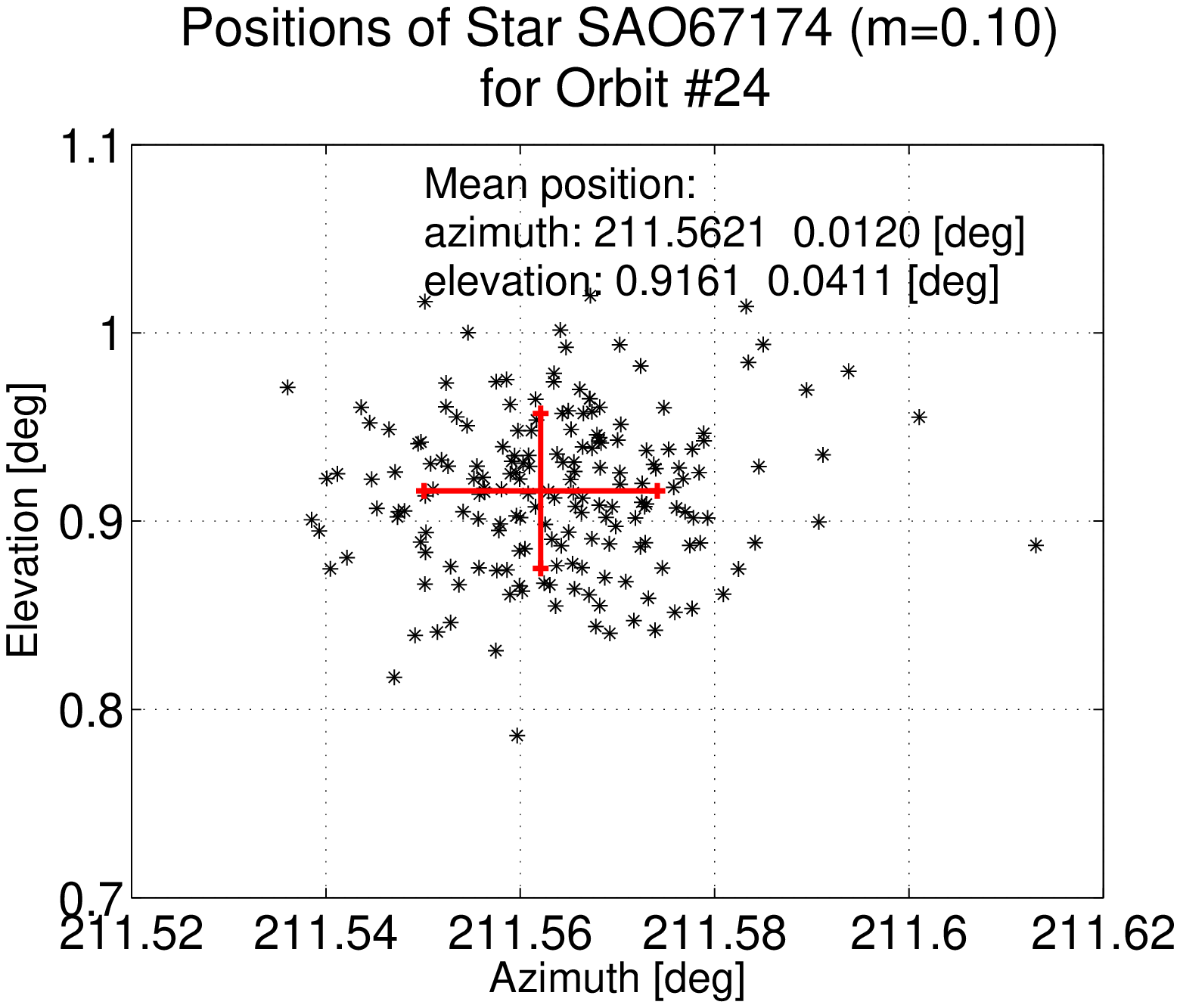}{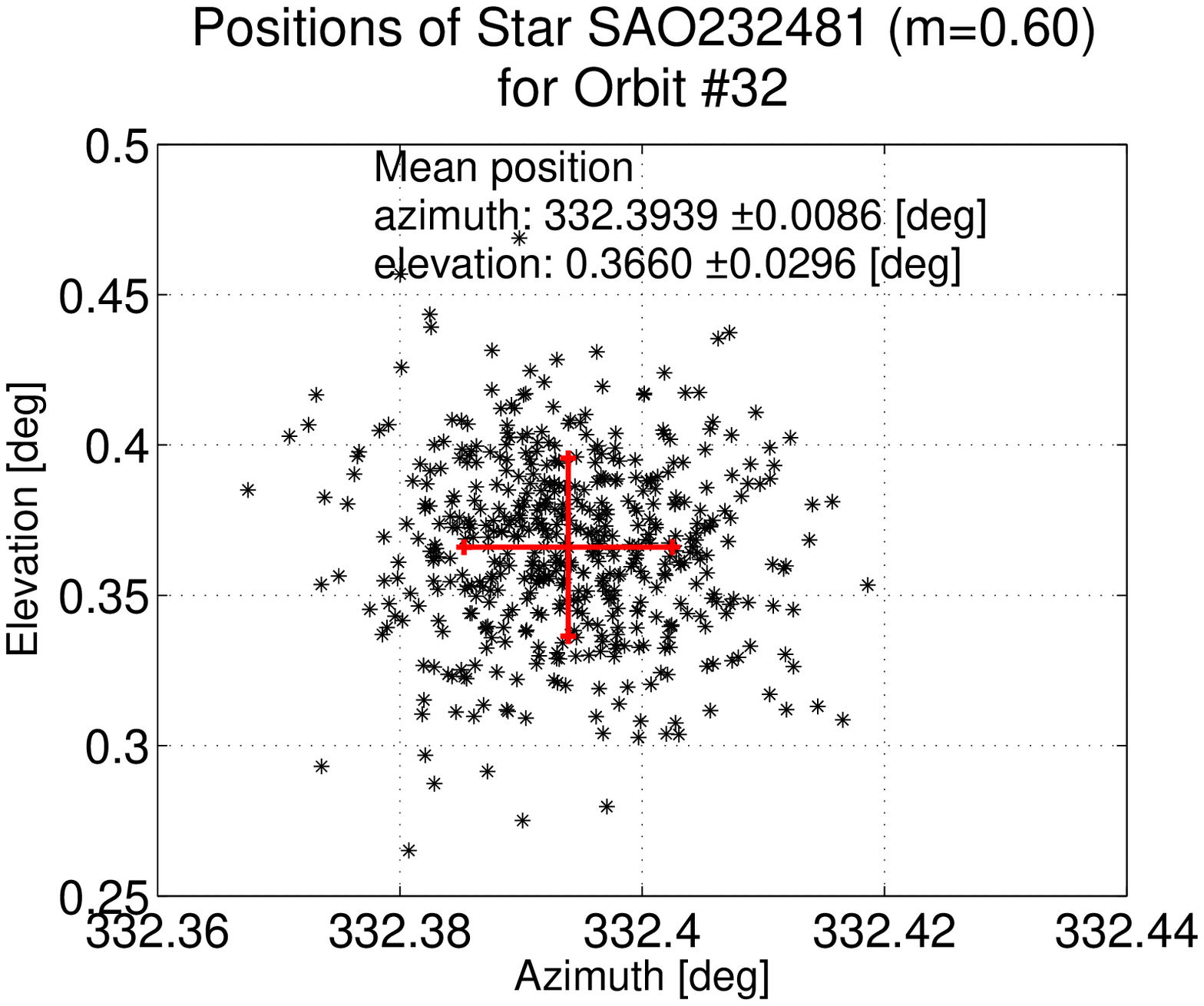}   
		\plottwo {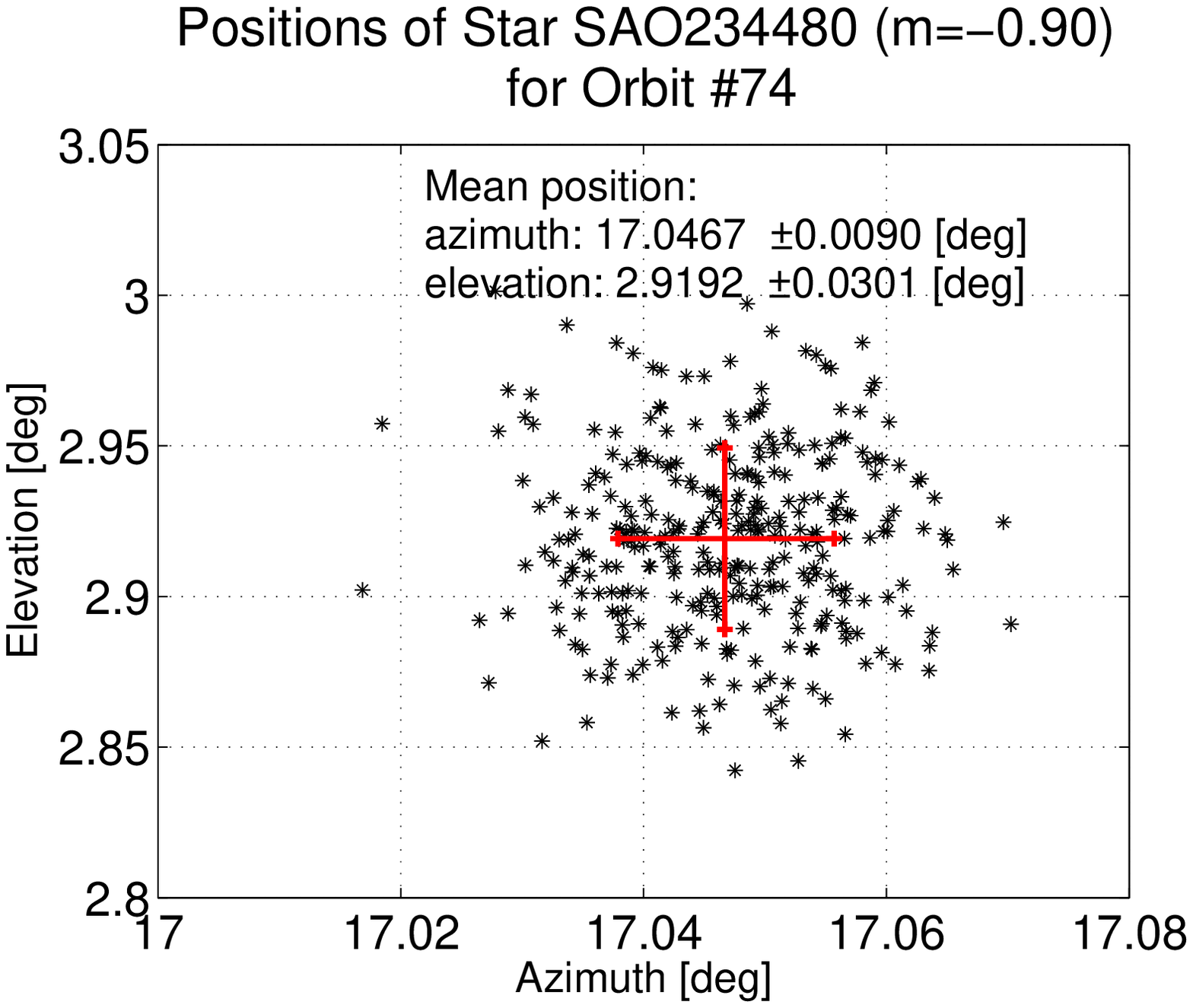}{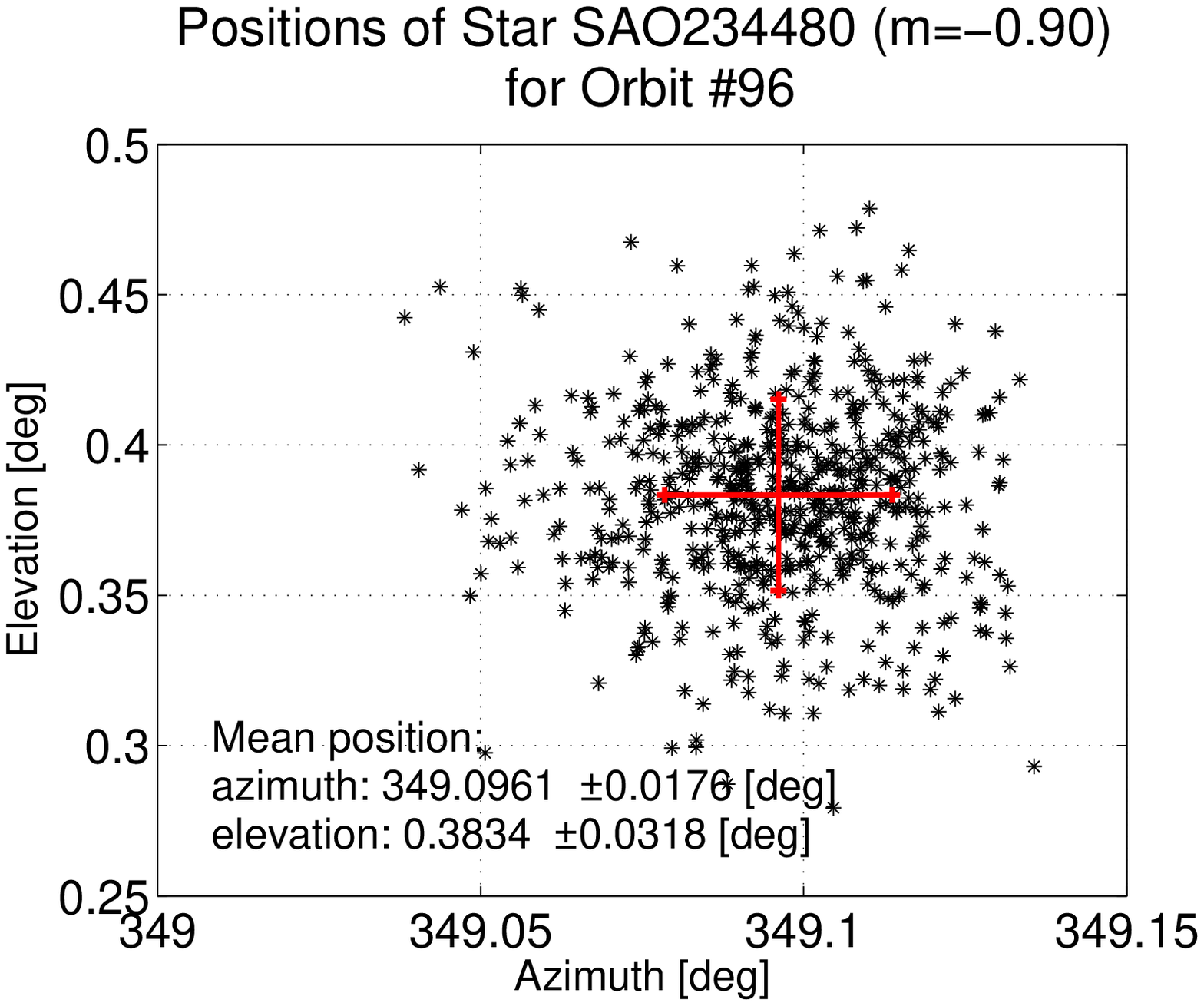}\\
  	\caption{Examples of star positions in the spacecraft coordinate system for selected orbits. Black dots mark a position obtained from a single histogram block, the red error bars indicate the mean values with standard deviations.}
 		\label{figSSStarSCPositn}
		\end{figure*}

As a first verification step for the IBEX-Lo boresight in the spacecraft system, we analyzed the positions of all identified stars in the spacecraft reference system for systematic deviations from the expected locations. For most of the suitable orbits, positions of two stars could be determined and the IBEX spin axis calculated. For a few orbits, the positions for either just one or as many as three stars could be determined. In total, 69 stars were processed. The positions of stars in the S/C reference system were retrieved separately from each suitable telemetry block. Examples of the determinations of these positions are shown in Fig. \ref{figSSStarSCPositn}. 	

\begin{figure*}[t]
		\centering
		\plottwo {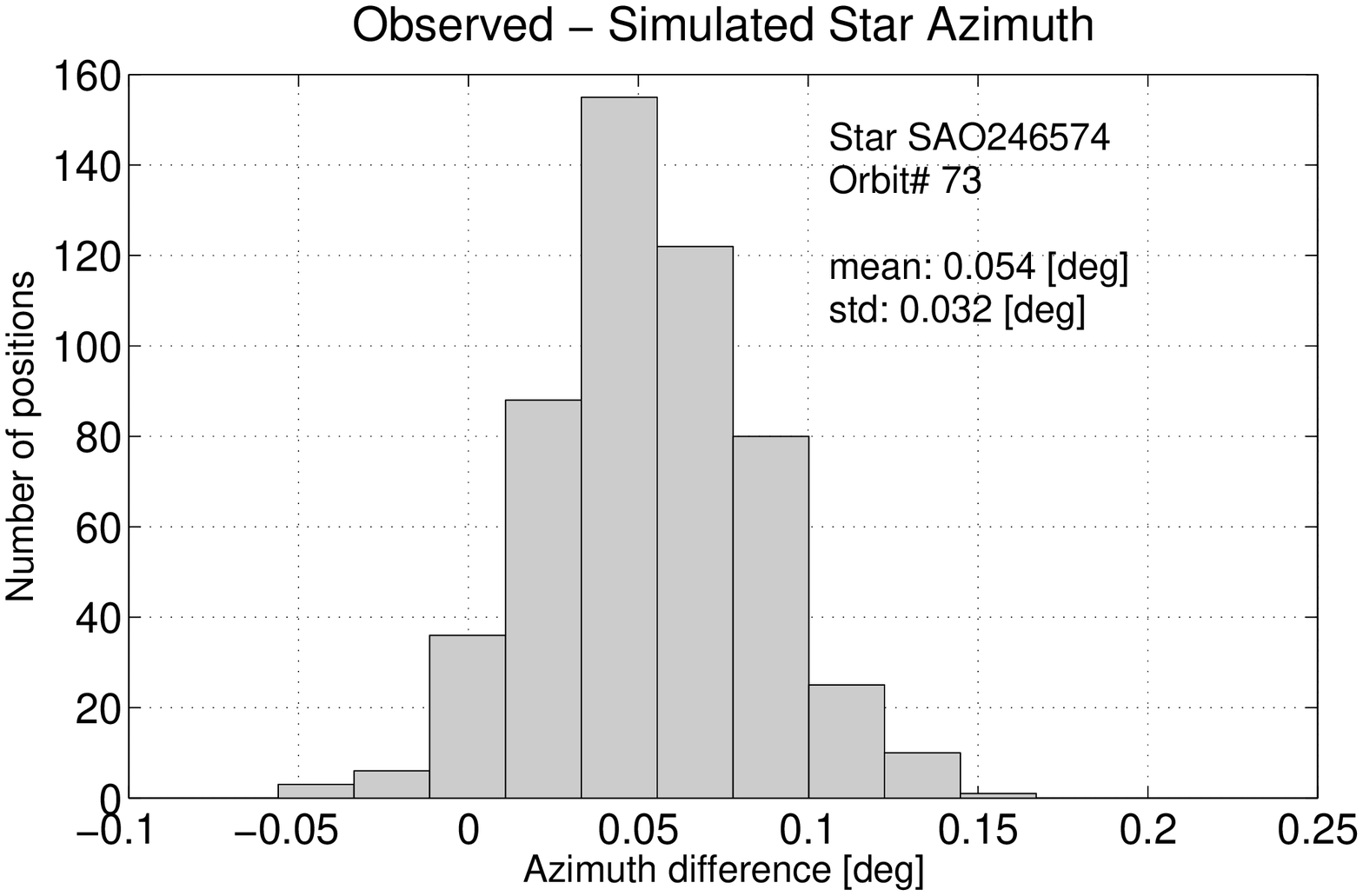}{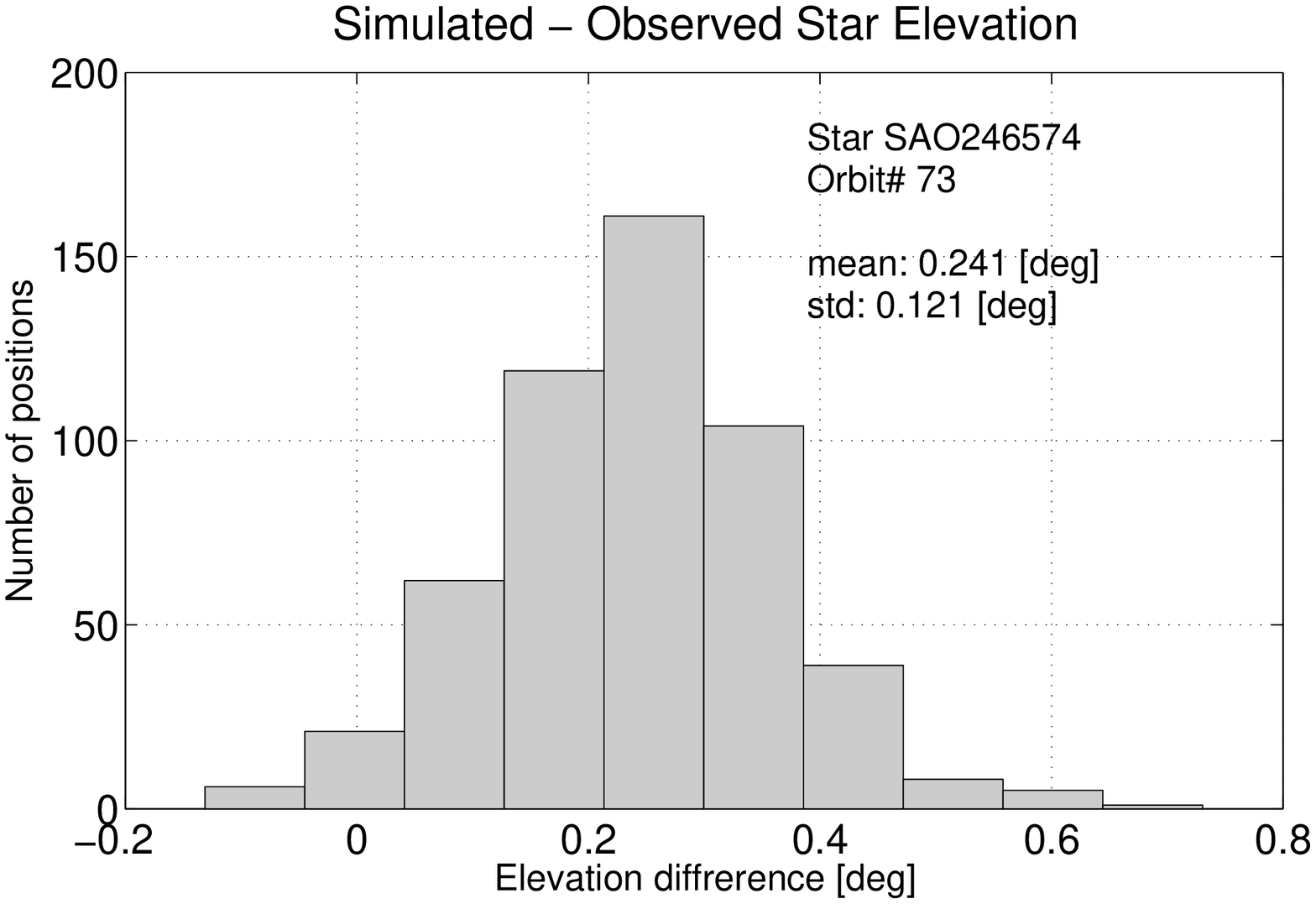}   
  	\caption{Histograms of differences between the simulated and observed azimuth (left-hand panel) and elevation (right-hand panel) of star SAO246574 in Orbit 73. The histograms illustrate the typical determination accuracy of star positions from Star Sensor data.}
 		\label{figSSPositFluctu}
		\end{figure*}

The positions of identified stars are somewhat distributed in both coordinates, with standard deviations varying from orbit to orbit. An example of statistical fluctuations of the differences between the simulated and actually measured azimuths and elevations for the star SAO246574 observed in Orbit 73 is shown in Fig. \ref{figSSPositFluctu}. 		
The positions were stable and consistent. The median value of standard deviations in the phase angle was 0.02\degr~and in elevation 0.06\degr. Such a behavior of a greater uncertainty in elevation than in spin angle was expected, as discussed in the previous section.  
 		
The angular separation between stars found for a given orbit is very stable. The distribution of angular separations for an example for orbit 73 has a standard deviation of less than $\sim0.04\degr$.		
		
\begin{figure*}[t]
		\centering
		\plottwo{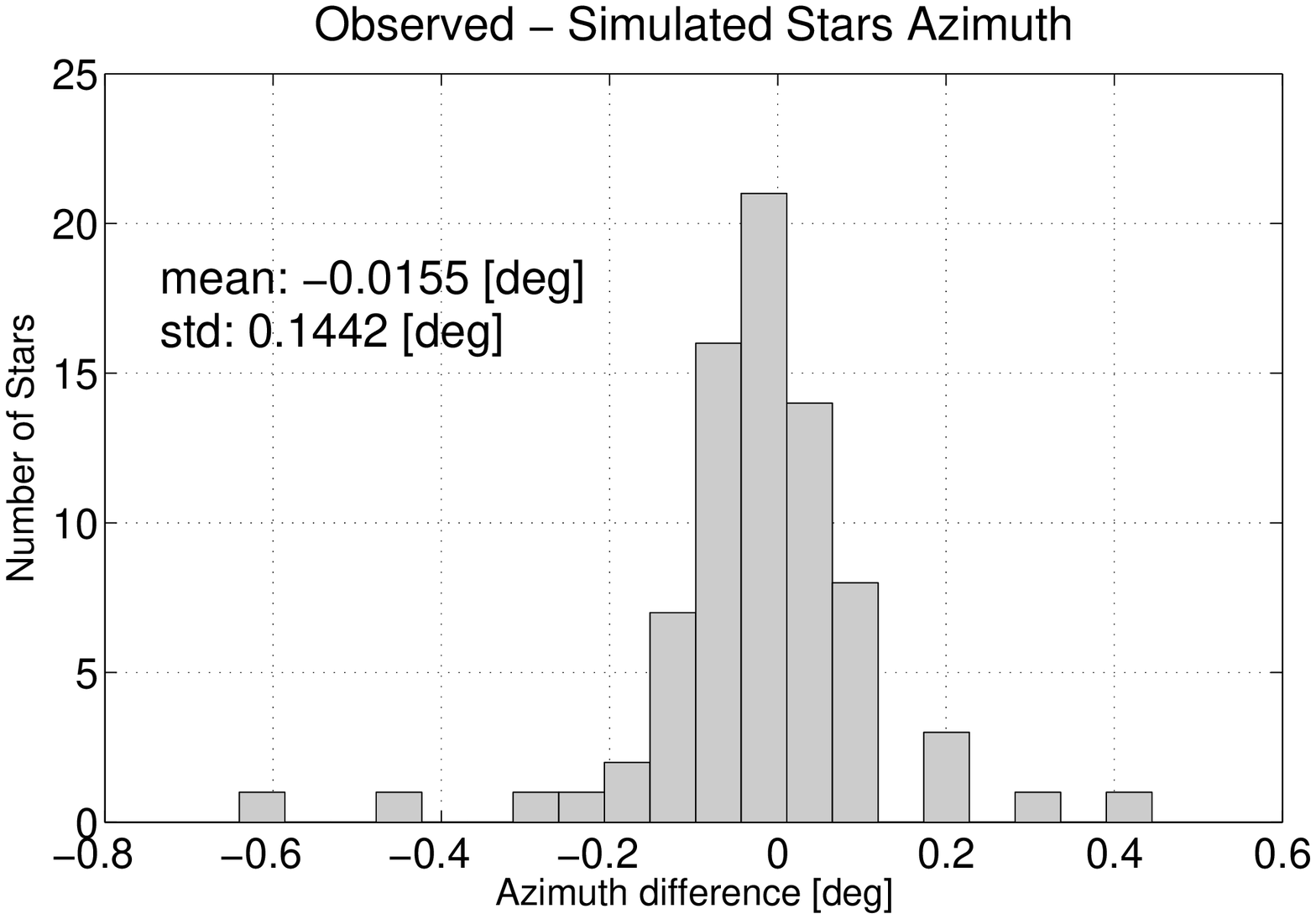}{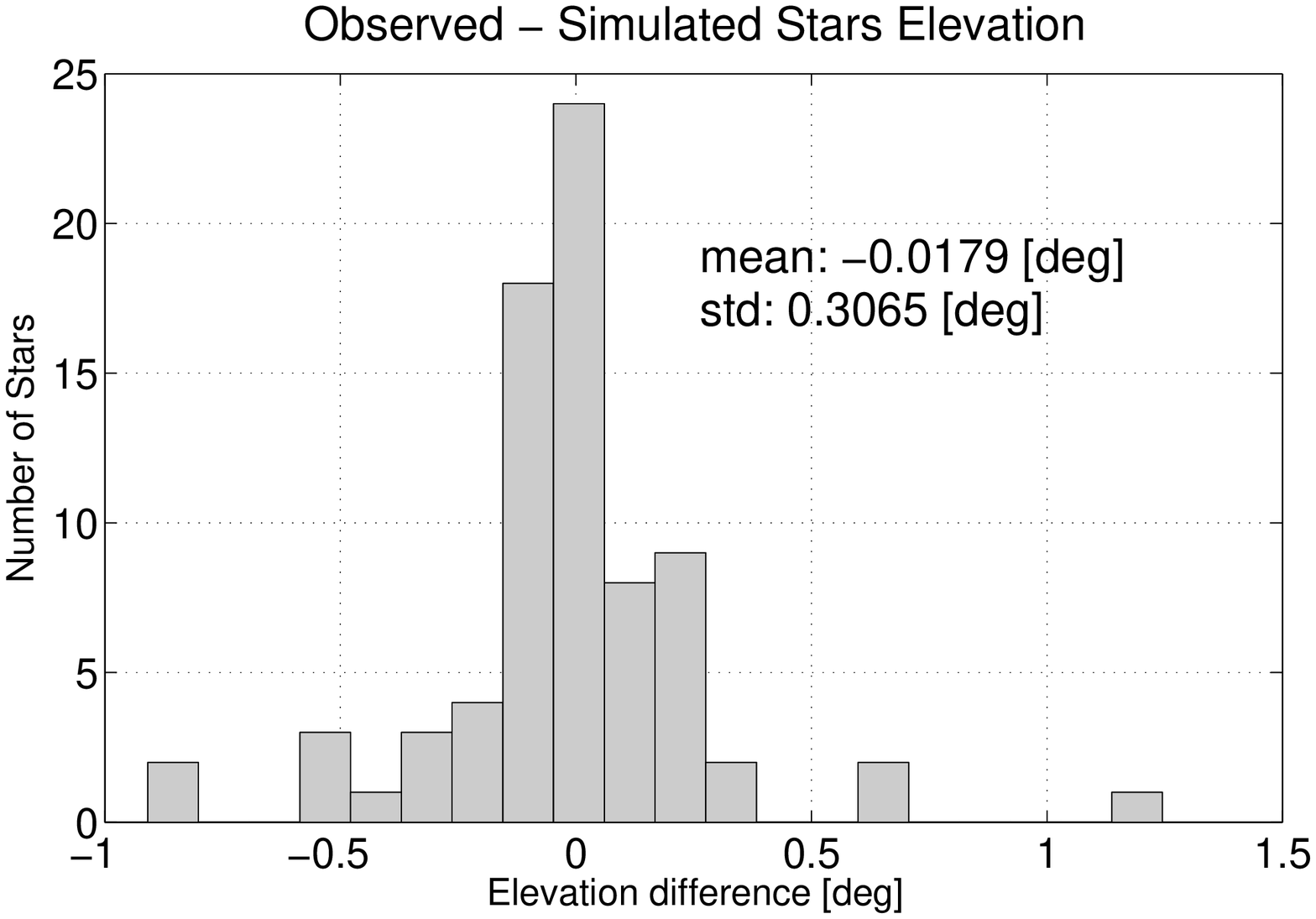}   
  	\caption{Histograms of the differences between the simulated and observed orbit-averaged azimuths (left-hand panel) and elevations (right-hand panel) of all stars observed by the Star Sensor.}
 		\label{figSSMeanDev}
		\end{figure*}

We calculated the mean positions of all identified stars, followed by the mean deviations of these positions from the simulated values based on the spin axis positions obtained from the IBEX Science Operations Center (ISOC) \cite{schwadron_etal:09a} using the spacecraft ACS. Histograms of the differences between the simulated and observed positions of the stars in the S/C reference system are shown in Fig. \ref{figSSMeanDev}. The uncertainty from the mean is given by $\sigma/\sqrt{N}$, where $N=69$ is the number of observed stars. Therefore, the mean deviation in the phase angle is equal to $-0.0155\degr \pm 0.017\degr$, while the mean deviation in elevation is $-0.0179\degr \pm 0.037\degr$. About 68\% of the results in elevation and about 85\% results in spin angle are within $\pm 0.2\degr$ from 0. We conclude that there is no statistically significant deviation of the Star Sensor boresight from the direction perpendicular to the ACS-determined spin axis to within 0.037\degr, nor in spin angle relative to the spin pulse to within 0.017\degr.
We conclude that there is no statistically significant deviation of the FoV from the Star Sensor and thus the IBEX-Lo pointing other than expected from the observed deflection of the spin angle from spin axis and from the ground-measured uncertainty in the IBEX-Lo mounting. In addition, the Star Sensor signal can be used to correct the spin phase information with high precision when bright objects, such as the Earth and Moon, blind the Star Tracker causing inaccuracies in the ACS data. 	
 		\subsection{Independent spin axis pointing determination using the Star Sensor }
 		
 		\begin{figure*}[t]
 		\epsscale{1.5}
 		\centering
 		\plottwo{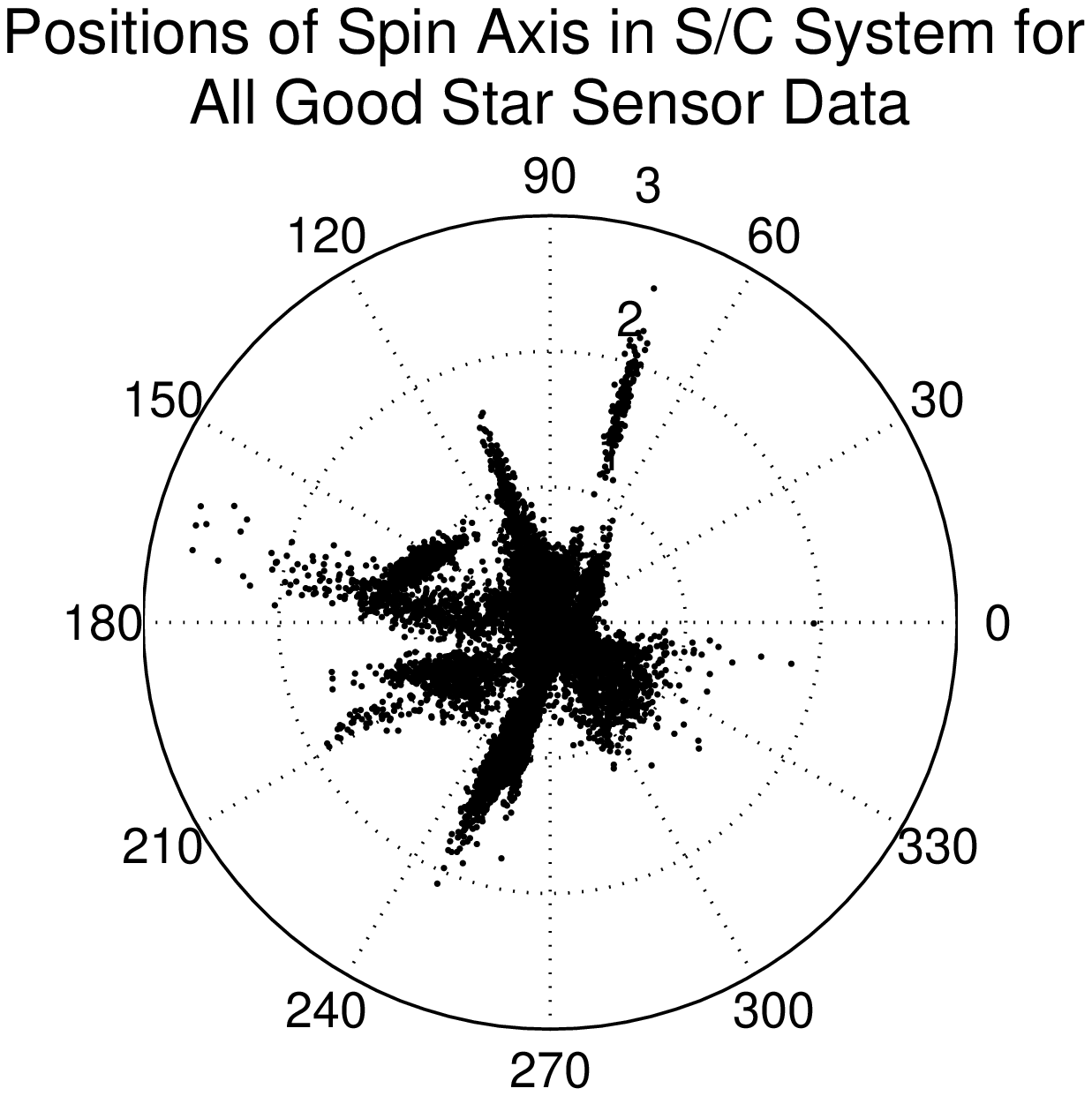}{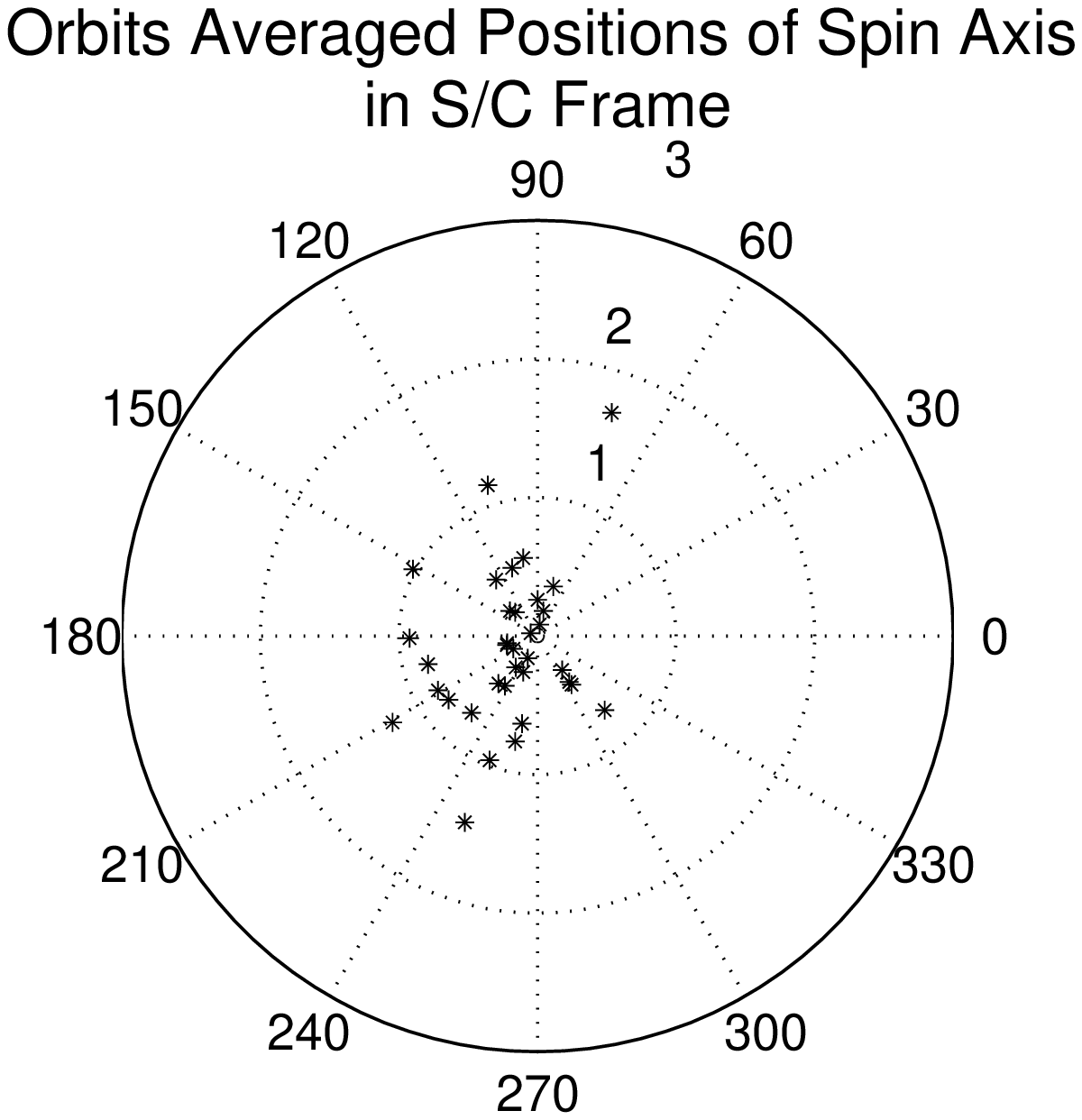}
  	\caption{Polar plots of the spin axis pointings in the spacecraft reference system determined from the observations of the Star Sensor. The left-hand panel presents separate determinations from all suitable spin blocks, the right-hand panel the positions averaged by orbits. The azimuth angles are marked at the circumference, the co-latitudes (polar distances) at the co-latitude rings. The z-axis (i.e. the center point) corresponds to the spin axis pointing determined by ISOC based on the Stars Tracker/ACS data.}
 		\label{figSSPolarPlots}
		\end{figure*} 
For all orbits with at least two clearly identified stars, we have calculated the pointing of the IBEX spin axis. We followed two paths: (1) we calculated the spin orientation separately for all spin blocks available from a given orbit, and (2) we calculated the spin axis from the mean star positions averaged over a given orbit. Subsequently, we calculated the deviation of our pointing from the ACS pointing. 			
The representation of the results of the spin axis determination from Star Sensor observations are presented in the S/C reference system in Fig. \ref{figSSAxisPerOrbit} in Appendix B. The S/C reference frame is oriented with the ACS-derived spin axis at the pole, and IBEX-Lo directed in the -Y direction. Typically, determinations for a given orbit form elongated streaks in the S/C reference system because the Star Sensor's accuracy for a star's elevation is lower than that of its azimuth. The quality of the IBEX pointing determinations varies between orbits because of the varying measurement conditions between orbits: relative position of the stars (differences in the azimuth and in the elevation), signal to background ratio of the stars, intensity and inhomogeneity of the background, presence of the stray light, and the finite accuracy of the projection of the stars azimuth on the Star Sensor 720-bin histogram. 		
		\begin{figure*}[t]
		\centering
		\epsscale{1.6}
		\plottwo {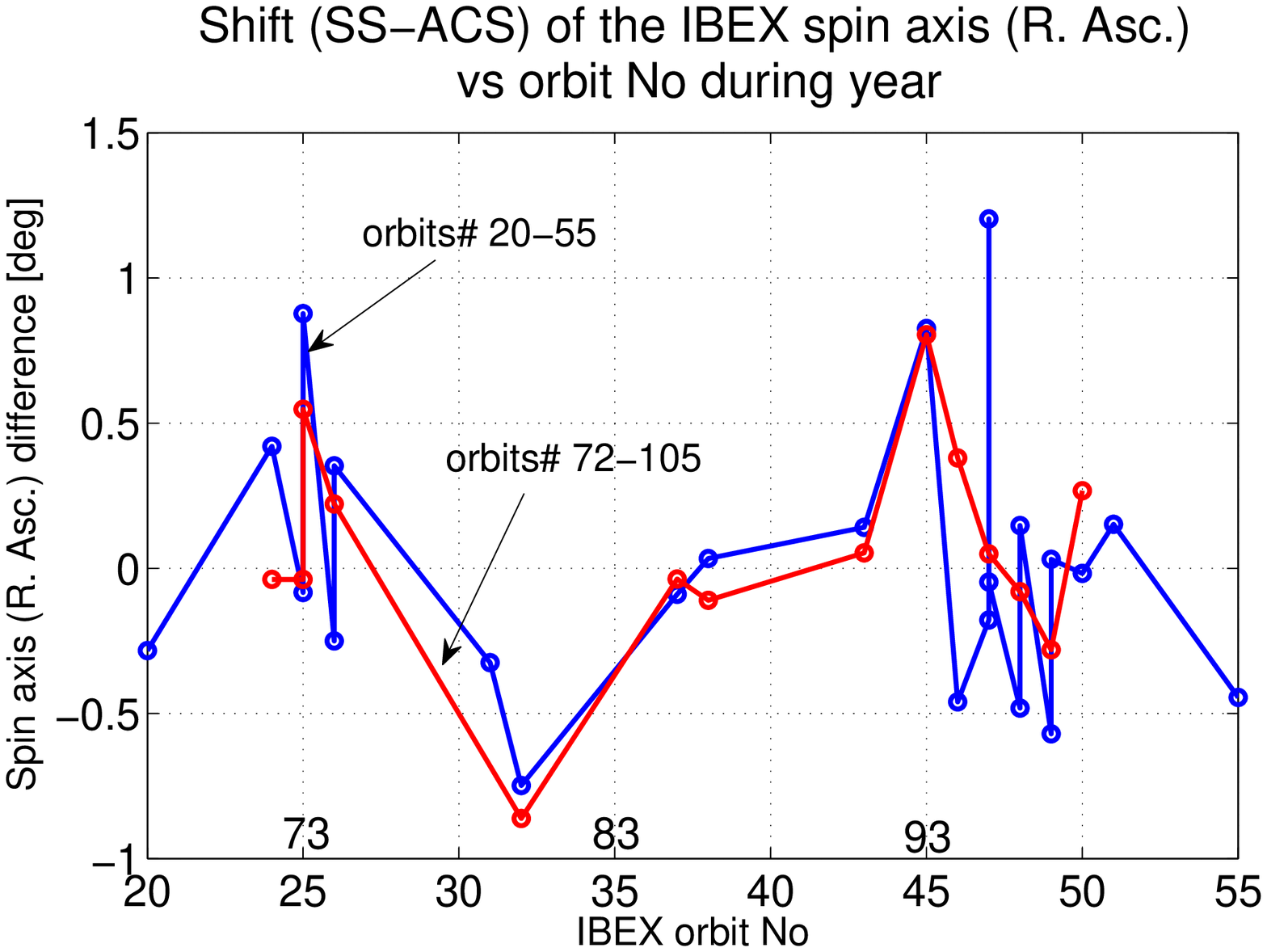}{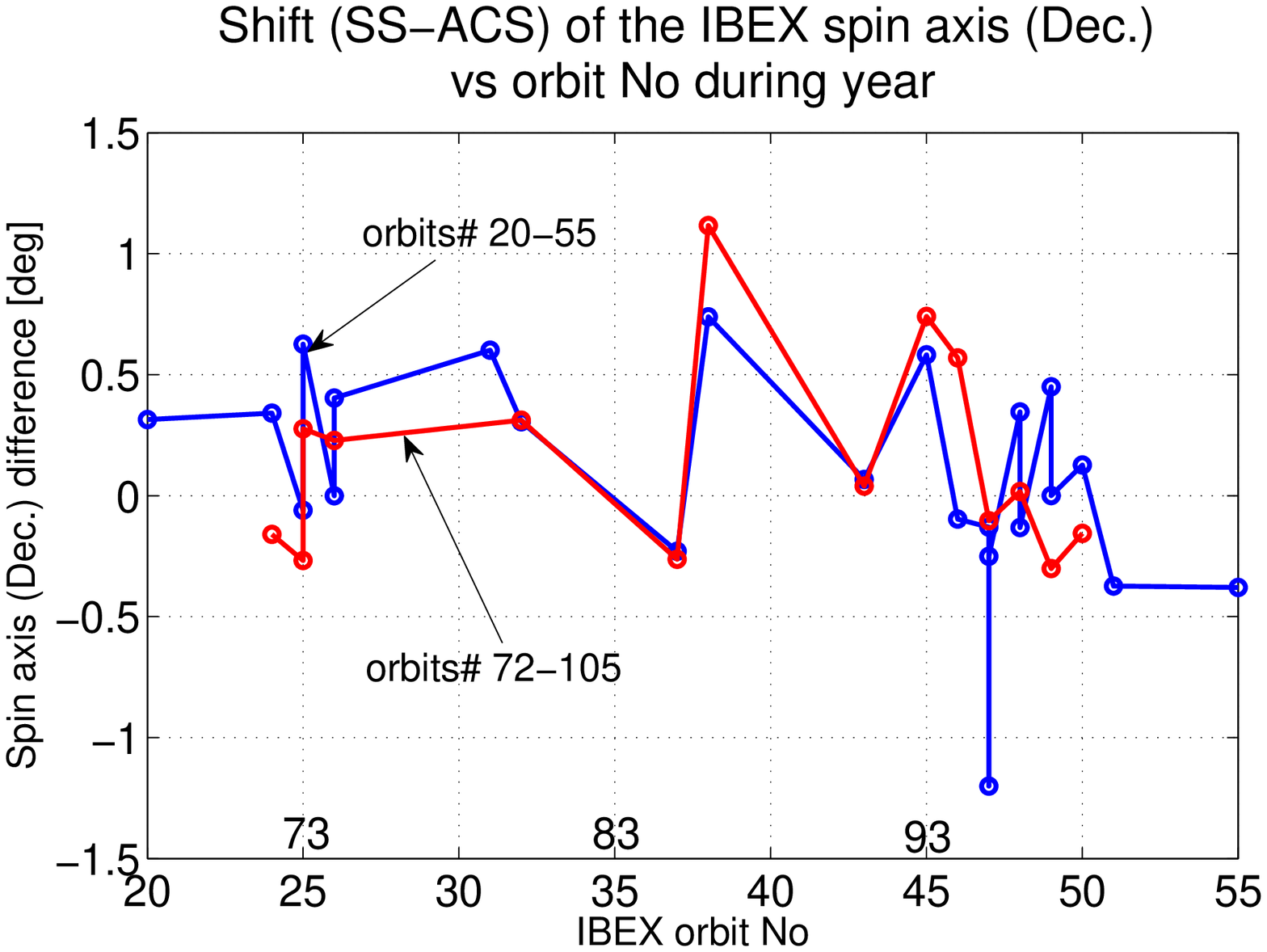}  
		{figSSSpinAxSeasonal01.png}{figSSSpinAxSeasonal02.png}\\
  	\caption{Seasonal variation of the differences between the Star Tracker/ACS and Star Sensor determinations of the orbit-averaged spin axis pointings in the equatorial reference system: rectascension (left-hand panel) and declination (right-hand panel). Blue lines mark the positions obtained for the orbits from 2009, red lines mark the positions obtained for the corresponding orbits from 2010.}
 		\label{figSSSpinAxSeasonal}
		\end{figure*} 
The positions of the spin axis as obtained from individual data blocks for all processed orbits are shown in the left panel of Fig. \ref{figSSPolarPlots} and the average positions for these orbits are shown in the right panel. The orbit averaged spin axis pointings determined by the Star Sensor are almost uniformly distributed around the pole. The small void region between 60\degr~and 330\degr~azimuth is due to the season during the year when no sufficiently bright star pairs can be observed. Importantly, almost all the center-lines through the uncertainty ellipses pass very close to the pole in the plot.

The spread between the ACS-derived spin axis and that derived by the Star Sensor is between 0.24\degr and 0.28\degr, which roughly agrees with the pre-launch measurements of the Star Sensor boresight relative to that of IBEX-Lo. From this, we conclude that there is no systematic statistically significant deviation of the IBEX-Lo boresight from the direction perpendicular to the ACS-derived spin axis. The spread obtained from the analysis of the Star Sensor data is due to the observing conditions and follows a yearly pattern caused by the positions of individual stars. This is illustrated in Fig. \ref{figSSSpinAxSeasonal} where differences in right ascension and declination of the spin axis between the ACS and Star Sensor determinations are shown as yearly time series based on orbit number. The accuracy of the spin axis pointing determined by the Star Sensor shows that it could be used for spacecraft attitude determination for almost all orbits in case of a Star Tracker failure, i.e., the Star Sensor indeed provides good redundancy to the spacecraft system. 

 	\section{Summary and conclusions}

We studied the accuracy of the Star Sensor pointing relative to the IBEX ACS to establish the absolute pointing of IBEX-Lo in astronomical coordinates during observations of neutral interstellar gas flow as presented by \cite{mobius_etal:12a, bzowski_etal:12a, bochsler_etal:12a, saul_etal:12a, lee_etal:12a}.  We analyzed IBEX ACS and Star Sensor data obtained during all orbits over the first two years of the mission, for which stars could be clearly resolved from background, including orbits when the neutral gas observations were carried out. 

There are four independent sources of uncertainty in the determination of the boresight of the IBEX-Lo sensor: 
\begin{enumerate}
\item Mechanical fabrication and mounting tolerances in the Star Sensor and IBEX-Lo lead to a 0.094\degr~azimuth (spin angle) uncertainty and a 0.102\degr~elevation angle uncertainty in the co-alignment of the Star Sensor and IBEX-Lo boresight. 
\item The spin axis is determined throughout each orbit from analysis of ACS attitude information is found to be stable to within 0.02\degr~in both directions.  
\item Observations of 69 stars and their positions were made using the Star Sensor (Fig. \ref{figSSMeanDev}), while the spacecraft ACS was used to determine the spacecraft spin axis. The mean deviation between simulated and observed stars in phase angle is equal to $-0.0155\degr \pm 0.017\degr$, while the mean deviation in elevation is $-0.0179\degr \pm 0.037\degr$. These observations show that, on average, there is no statistically significant deviation of the Star Sensor boresight from the direction perpendicular to the ACS-determined spin axis to within 0.037\degr, nor in spin angle relative to the spin pulse to within 0.017\degr.
\item Both the Star Sensor and ACS measurements are affected by a 1~ms spin period granularity, which leads to up to a 0.025\degr~uncertainty in spin-phase.
\end{enumerate}
	\begin{table*}
	\begin{center}
	
	\caption{Measurements uncertainties in the IBEX-Lo pointing based on Star Sensor and ACS \label{tabMeasureUncert}}
	\begin{tabular}{c l l l }
	\tableline\tableline
	 & Uncertainty Type & Spin Angle & Elevation Angle \\ 
	 & &Uncertainty (deg) & Uncertainty (deg) \\ \tableline
	1) & Star Sensor Mounting Knowledge & 0.094 & 0.102 \\ \tableline
	2) & Spin axis stability & 0.02 & 0.02 \\ \tableline	
	3) & ACS-determined spin-axis & 0.017 & 0.037 \\ \tableline	
	4) & Spin Period granularity & $<$0.025 &  -\\ \tableline
	& Uncertainty in IBEX-Lo Attitude & 0.099 & 0.104 \\ 
	& using Star Sensor & & \\ \tableline
	&	Uncertainty in IBEX-Lo Attitude & 0.101 & 0.110	\\
	& using ACS & & \\ \tableline
	\end{tabular}
 	\end{center}
	\end{table*}


Table \ref{tabMeasureUncert} lists each of these measurement uncertainties and combines them to yield spin angle and elevation angle uncertainties in the IBEX-Lo pointing based on instantaneous Star Sensor and ACS attitude determination. Uncertainty in co-alignment of the Star Sensor and IBEX-Lo boresights dominates attitude knowledge using both the Star Sensor and the ACS. We conclude that no systematic correction for boresight geometry needs to be introduced in the analysis of neutral interstellar gas flow observed by IBEX-Lo and the net uncertainty in the instantaneous IBEX-Lo pointing is $\sim 0.1\degr$ in both spin angle and elevation.

Our analysis shows that Star Sensor information can be used for spacecraft attitude corrections during periods when the spacecraft Star Tracker (used by the ACS) is disabled by bright objects in its field-of-view. In addition, we have demonstrated that Star Sensor measurements provide independent determination of the spacecraft spin axis during most of the orbits with slightly larger uncertainties than the ACS-determined spin axis.  Thus, the Star Sensor can be used for spacecraft attitude determination and provides redundancy for the Star Tracker in case of any problems.
 	
 	\acknowledgments
Acknowledgments: The use of the SPICE platform from the NASA's Navigation and Ancillary Information Facility (NAIF) at the NASA Jet Propulsion Laboratory and of the Hierarchical Equal Area isoLatitude Pixelisation (HEALPix) software as well as of the digitized SAO Star Catalog are gratefully acknowledged. M.H. and M.B. were supported by the Polish Ministry for Science and Higher Education grants N522 002 31/0902 and NS-1260-11-09. This work was supported by the IBEX mission as a part of NASA's Explorer Program.

\bibliographystyle{apj}
\bibliography{iplbib}{}

 \appendix 
 \section{(Appendix) Star Sensor Simulations Program}

The Star Sensor is able to register bright stars, the naked-eye outer planets (Mars, Jupiter and Saturn) and the Moon. All of them can be used to determine the pointing of the spin axis and all of them can be simulated by the Star Sensor Simulations Program, which is needed to identify the objects observed by the Star Sensor with the objects in the reference star catalog and to provide insight into the sky portion observed. 

The Star Sensor Simulation Program must identify a list of objects that will enter the FoV for a given orientation of the IBEX spin axis at a given time. It is also desirable to determine the objects located in the adjacent regions of the sky in the case the true rotation axis differs from the assumed one and the observed signal includes the stars that basically should not be visible. Determination of positions of these objects on the sky and verification if they are within the visible sky strip at the given moment is done with the use of specially developed procedures that use the SPICE software platform \citep{acton:96a} from the NASA Navigation and Ancillary Information Facility. 

To determine which stars will be visible for the Star Sensor in a given orbit for a given spin axis pointing one must convert the positions of the bright stars from a reference star catalog to the spacecraft frame. From the many star catalog systems available for attitude determination, we chose the Smithsonian Astrophysical Observatory Catalogue (SAO FK5). The SAO catalog contains almost 260 000 stars down to about 10.0 visual magnitude. The magnitudes are available only as photographic and/or photo-visual values with an accuracy of $\sim 0.5$ magnitude.  The uncertainty of positions is $\sim 0.5$ arc-sec \citep{wertz:78}. Because the model of the sky background used in the Simulation Program covers the stars up to the magnitude $m_v = 6.5$, we actually used a subset of the SAO catalog that includes only the stars brighter than $m_v = 6.5$. This was done for practical reasons: Star Sensor is not able to resolve weaker stars (actually, the sensitivity limit for meaningful applications is about $m = 3.5$) and limiting the source catalog to fewer entries speeds up the process of matching the catalog stars with the observed ones.

		\begin{figure}[t]
		\epsscale{0.6}
		\centering
		\plotone{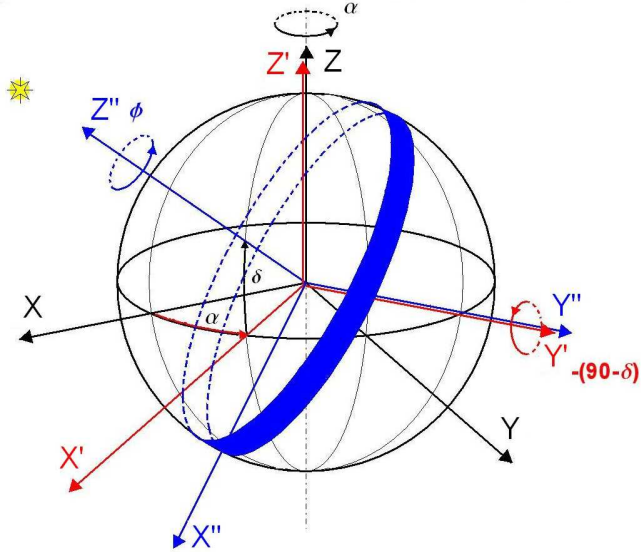}   
  	\caption{ Illustration of the sky strip scanned by the Star Sensor during the IBEX rotation. The strip is marked in blue; the blue Z'' axis marks the rotation axis of the spacecraft, which is also the Z-axis of the spacecraft reference system. The black-red Z axis marks the pole of the equatorial reference system, in which the stars position are listed in the catalog. The rotation axis is within $~1\degr$ of the ecliptic plane.}
 		\label{figSSReferenceSystems}
		\end{figure}

The Star Sensor is able to observe objects located in a strip of the sky between $+3.5\degr$ and $-5.0\degr$ from the x-y plane of the IBEX spacecraft, which is perpendicular to the rotation axis (see Fig. \ref{figSSReferenceSystems}). The rotation axis is maintained within a few degrees from the Sun. Thus the field of view of the Sensor changes from orbit to orbit. Hence for each simulation for a given date and given pointing of the spin axis the sub-catalog is updated for the date of observations and converted from the equatorial coordinate system to the S/C coordinate system. The new catalog makes basis for the Star Sensor output signal simulation and scanned stars identification. 
			
Since the Star Sensor is able to resolve the three outer naked-eye planets, the catalog of potentially visible stars that is constructed from the subset of the SAO catalog is supplemented with the calculated positions of these three planets valid for the time of the simulation. Since the planets are practically point objects and their proper motion on sky during the $\sim16$ minutes of the duration of one block of Star Sensor observations is practically 0, they are treated as stars and included into the Potentially Visible Objects Catalog constructed for each simulation. 
		\begin{figure}[t]
		\epsscale{0.6}
		\centering
		\plotone{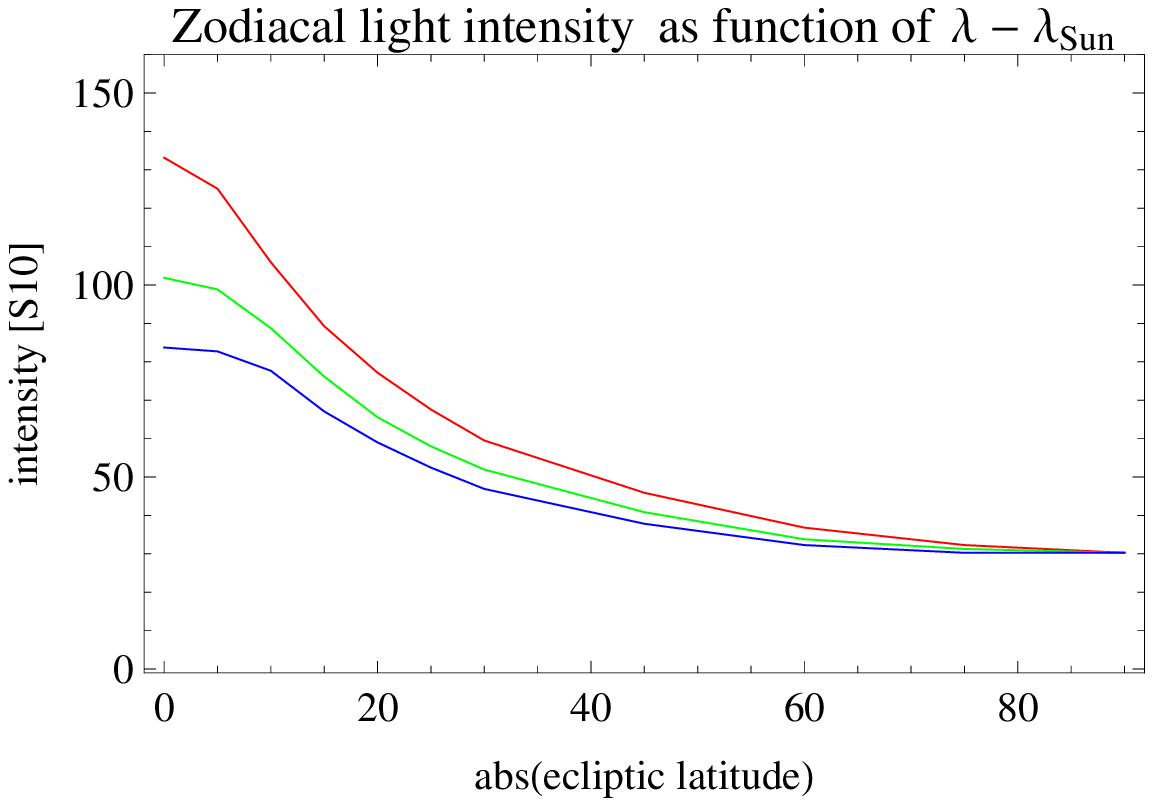}  
  	\caption{Intensity of the zodiacal light in the S10 units as function of ecliptic latitude for 3 selected $\left|\lambda - \lambda_{\mathrm{Sun}}\right|$: $75\degr$ (red), $90\degr$ (green), and $105\degr$ (blue), adopted from \cite{leinert_etal:98a} and used in the Star Sensor Simulation Program.}
 		\label{figSSZodLight}
		\end{figure}
All point-like objects are always observed against a diffuse sky background and since the Star Sensor is a pin-hole camera, not an imaging device, the signal registered is a convolved sum of contributions from the point sources and the background. The components of the background are the diffuse Galactic and inter-galactic light, faint unresolved stars, and the zodiacal light. The first two components of the background are practically invariable in time, but the zodiacal light intensity is a function of the angular distance of the line of sight from the Sun and thus varies during one orbit. An extensive review of the sky background in the optical domain and its components is provided by \cite{leinert_etal:98a}. 
		
The zodiacal light comes up mostly because of the scattering of sunlight off the interplanetary dust grains. Interplanetary dust is distributed symmetrically around the ecliptic plane, so the intensity as seen from the Earth orbit is also symmetric around the ecliptic, and because of the optics of the scattering, the intensity is a function of the absolute value of difference between the ecliptic longitude of the line of sight and the Sun. In the numerical model implemented for the Star Sensor Simulations Program we adopted the model of intensity of the zodiacal light from \cite{leinert_etal:98a}, Table 16, with a correction defined in their Eq.~(23).
The intensity of the zodiacal light is shown in Fig. \ref{figSSZodLight}. The FoV of the Star Sensor occupies a $360\degr$ band perpendicular to the ecliptic, but since the spin axis of the spacecraft is always within a few degrees off the Sun, the strip of the sky visible to the Star Sensor covers a very limited band of ecliptic longitudes that differ from the longitude of the Sun by about $90\degr$. Hence the model of the zodiacal light adopted in the Star Sensor Simulation Program could be based on a limited range of longitudes differences $75\degr < \left|\lambda - \lambda_{\mathrm{Sun}}\right| < 105\degr$. 

		\begin{figure*}[t]
		\centering
		\plotone{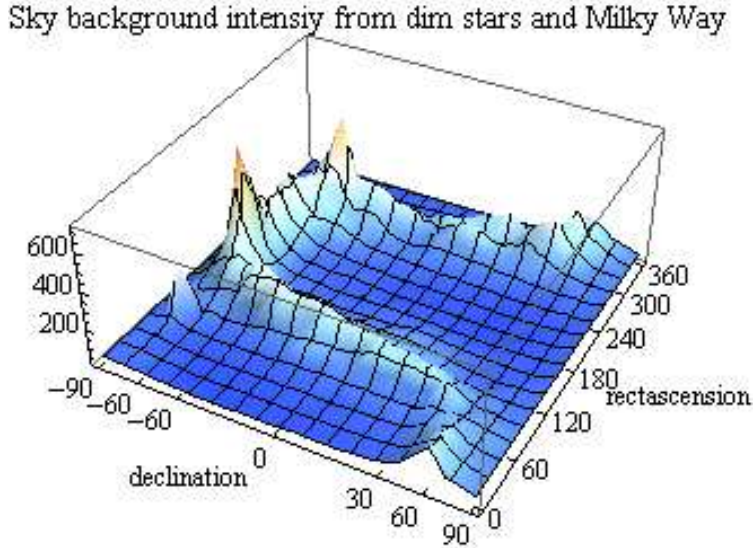}   
  	\caption{Intensity in the S10$_{\mathrm{Sun}}$ units of the diffuse Galactic and starlight as adopted in the model of the sky background used in the Star Sensor Simulation Program, shown as function of rectascension and declination.}
 		\label{figSSStarLight}
		\end{figure*}
			
Another component of the model of the sky background is the diffuse Galactic light. It was again adopted after \cite{leinert_etal:98a}, based on measurements of the sky background from the Pioneer spacecraft performed beyond the region occupied by most of the interplanetary dust. We used the data from Table 37 in \cite{leinert_etal:98a}, which are in the S$10_{\mathrm{Sun}}$ units. Since the spectrum of the zodiacal light is very similar to a slightly reddened solar spectrum, the models of the zodiacal light and of the starlight were compatible and since the spectral range of the Star Sensor PMT corresponds to the visible spectrum, it reflected relatively well the sensitivity. The data were on a $10\degr$ by $10\degr$ grid in the equatorial coordinates and covered stars up to the visual magnitude $m = 6.5$. Hence all the brighter stars, even though they could not be clearly resolved by the Star Sensor, had to be included in the background model directly from the star catalog. A plot of the background from the Galactic diffuse light and unresolved starlight is shown in Fig. \ref{figSSStarLight}.

		\begin{figure*}[t]
		\epsscale{0.9}
		\centering
		\plotone{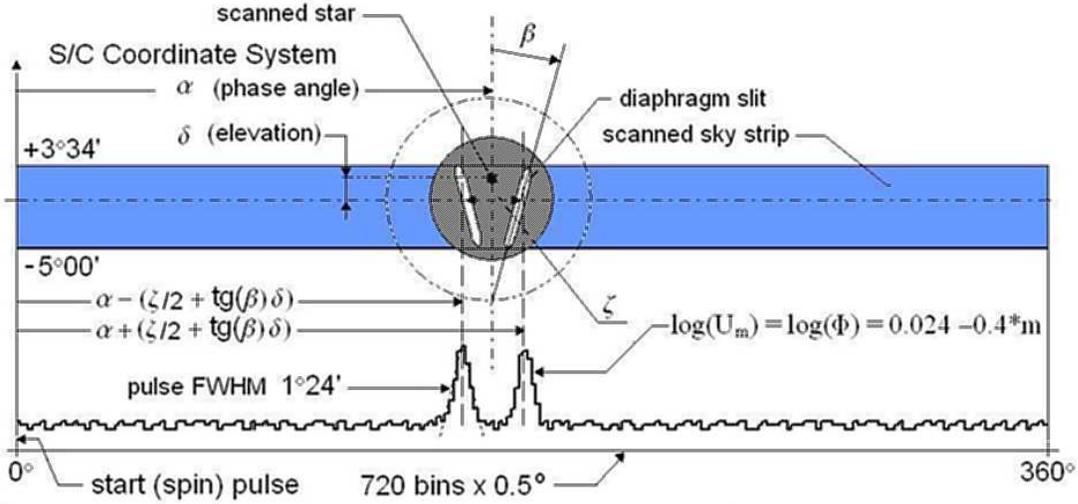} 
  	\caption{Schematic view of scanning of the sky by the Star Sensor. The strip of the sky visible to the Star Sensor is marked in blue. Shown is one bright star within this Visible Strip (black dots), being scanned by the split-V aperture of the Star Sensor. Below the Visible Strip the simulated signal is shown. The star being scanned makes the two peaks of the output voltage.}
 		\label{figSSscanSchem}
		\end{figure*}
		
The simulated signal of the Star Sensor from a point source is formed as illustrated in Fig. \ref{figSSscanSchem}. Shown is the Star Sensor aperture scanning the Visible Strip of the sky. The signal from a point source in the absence of the background is shown below the aperture as a double pulse. The output pulses can be approximated as triangles with the FWHM determined by geometrical dimensions of the collimator. The relation between the star magnitude $m$ and the output voltage $U$ (shown in the figure) is logarithmic. The Star Sensor Simulations Program simulates also the random component in the flux, calculating it as a Gaussian noise component of the voltage. The relations between the positions of the peaks and their angular separation and the coordinates of the scanned star in the S/C reference system are repeated in the figure after Eqs. (\ref{eqSSStarAzim}, \ref{eqSSStarElev}).
The signal observed is a superposition of the signals from all sources. In order to simulate the diffuse background and the stars, we divided the area of the Visible Strip into $\sim 800\,000$ pixels, adopting a tessellation suggested by \cite{gorski_etal:05a} in their Hierarchical Equal Area Latitude Pixelization (HEALPix) system, mapped at the spacecraft coordinates. The equal-area pixels are approximately rectangular in shape and arranged in latitudinal rings; the angular distances between the centers of the neighboring pixels are equal to each other. These features of this sky tessellation system made it well suitable for our purposes. We mapped (separately) the models of the zodiacal light and of the diffuse starlight for a given time and spin axis pointing at the HEALPix pixels and then calculated the brightness of each pixel. The model signal from the diffuse background is a superposition of the signals from virtual point sources, located at the centers of the pixels. The signals from the catalog stars and planets (if present in the FoV) are then added to form the output model signal. An example of the fidelity of the model signal as compared with the actual data is shown in Fig. \ref{figSSstarIdent}.		
		
In anticipation of an insufficient number of well resolved stars against background, we prepared for determination of the spin axis pointing from observations of the Moon. The Moon crosses the FoV approximately once per fortnight and it is so bright that the Star Sensor has to be switched into a special Moon Mode to prevent the signal from being saturated. The voltage on the PMT in the Moon Mode is reduced, but at a cost of reducing the signals from stars below the detectability threshold. The Moon is a fast-moving object and its angular size at IBEX locations varies from $~0.25\degr$ to $~4\degr$. Since it is always observed close to the half-moon phase, the Simulation Program must take into account its characteristic shape and the fact that the center of light is offset from the center of mass position on sky. The Simulation Program projects the expected shape of the Moon body at the HEALPix pixels and is able to simulate a realistic Moon signal, but because of the very complex nature of the signal and its interpretation on one hand and because point objects turned out to be sufficient to fulfill the task of the Star Sensor on the other hand, this mode of operation of the Star Sensor Simulations Program has never been exercised to interpret the Star Sensor observations.

 \section{(APPENDIX) Spin axis pointing in the spacecraft frame for individual orbits}

\begin{figure}[t]
		\begin{tabular}{ccc}
		\centering
		\includegraphics[scale=0.4, trim = 3mm 0mm 0mm 0mm, 
clip]{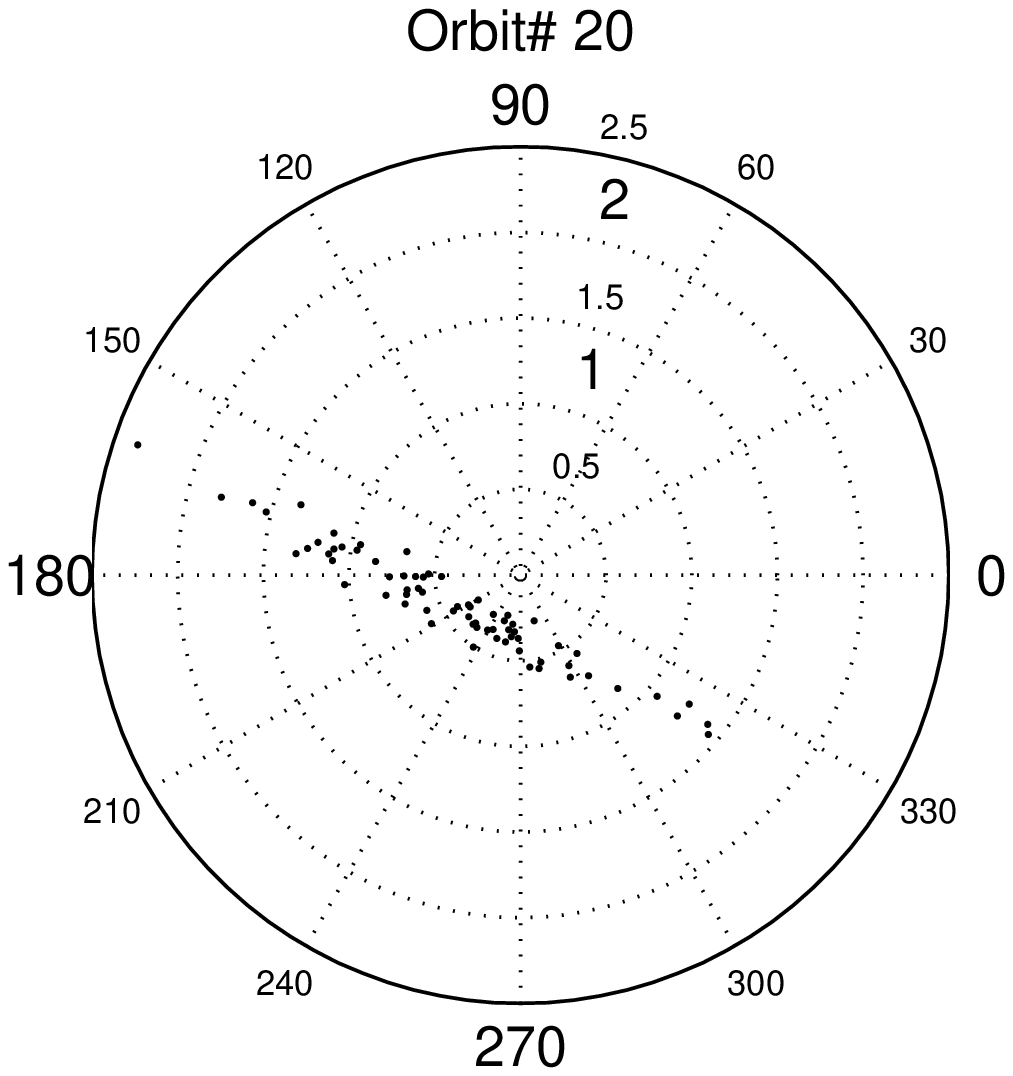}&\includegraphics[scale=0.4, trim = 3mm 0mm 0mm 0mm,
clip]{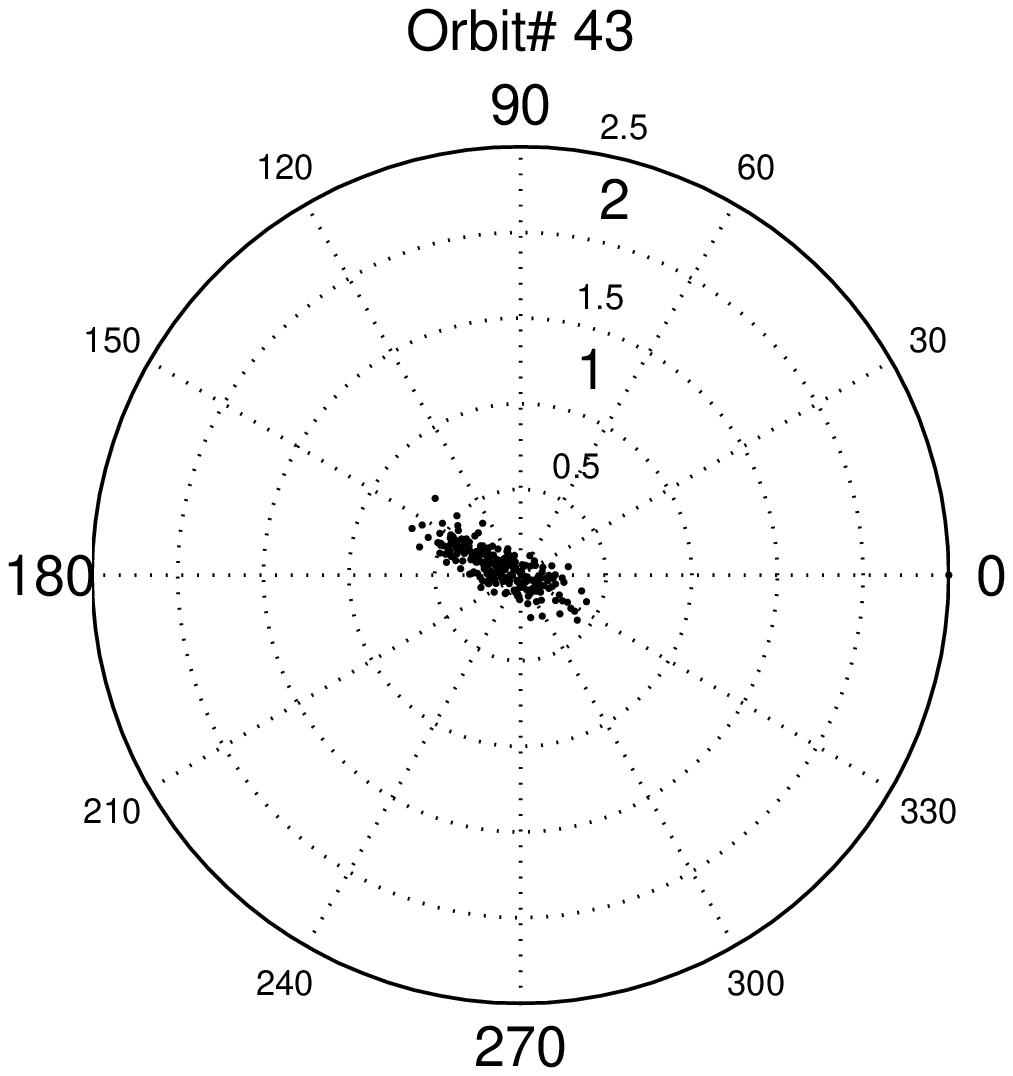}&\includegraphics[scale=0.4, trim = 3mm 0mm 0mm 0mm, 
  clip]{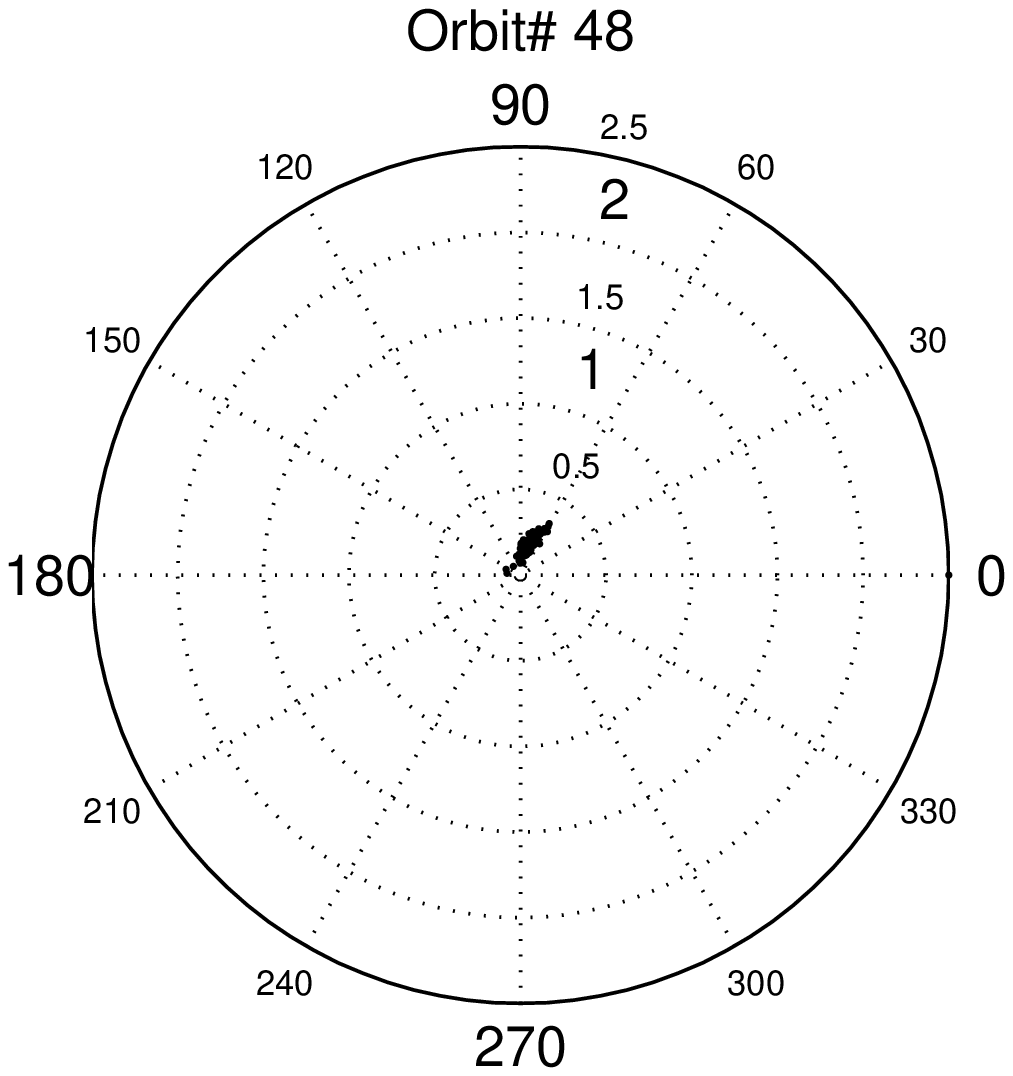}\\
		\includegraphics[scale=0.4, trim = 3mm 0mm 0mm 0mm, 
clip]{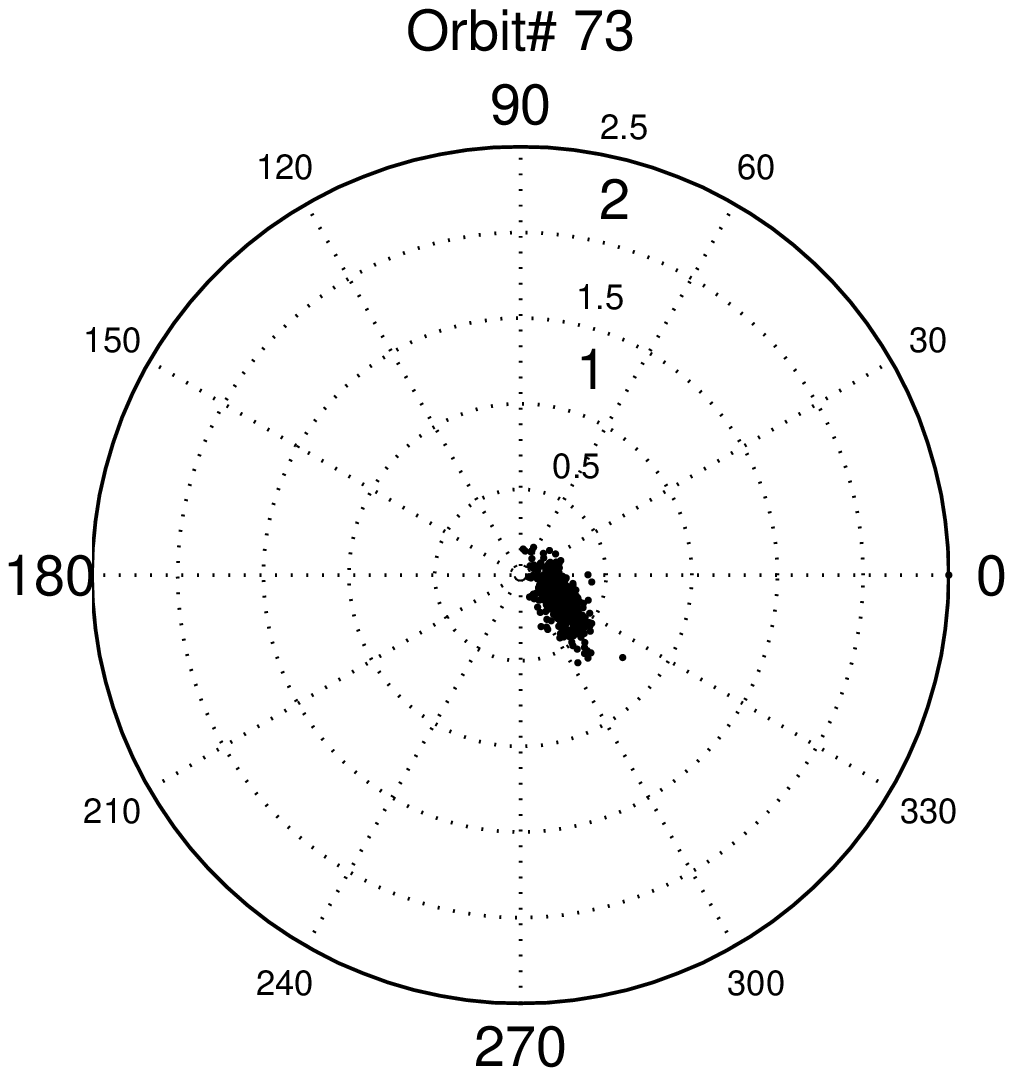}&\includegraphics[scale=0.4, trim = 3mm 0mm 0mm 0mm,		
		 	 clip]{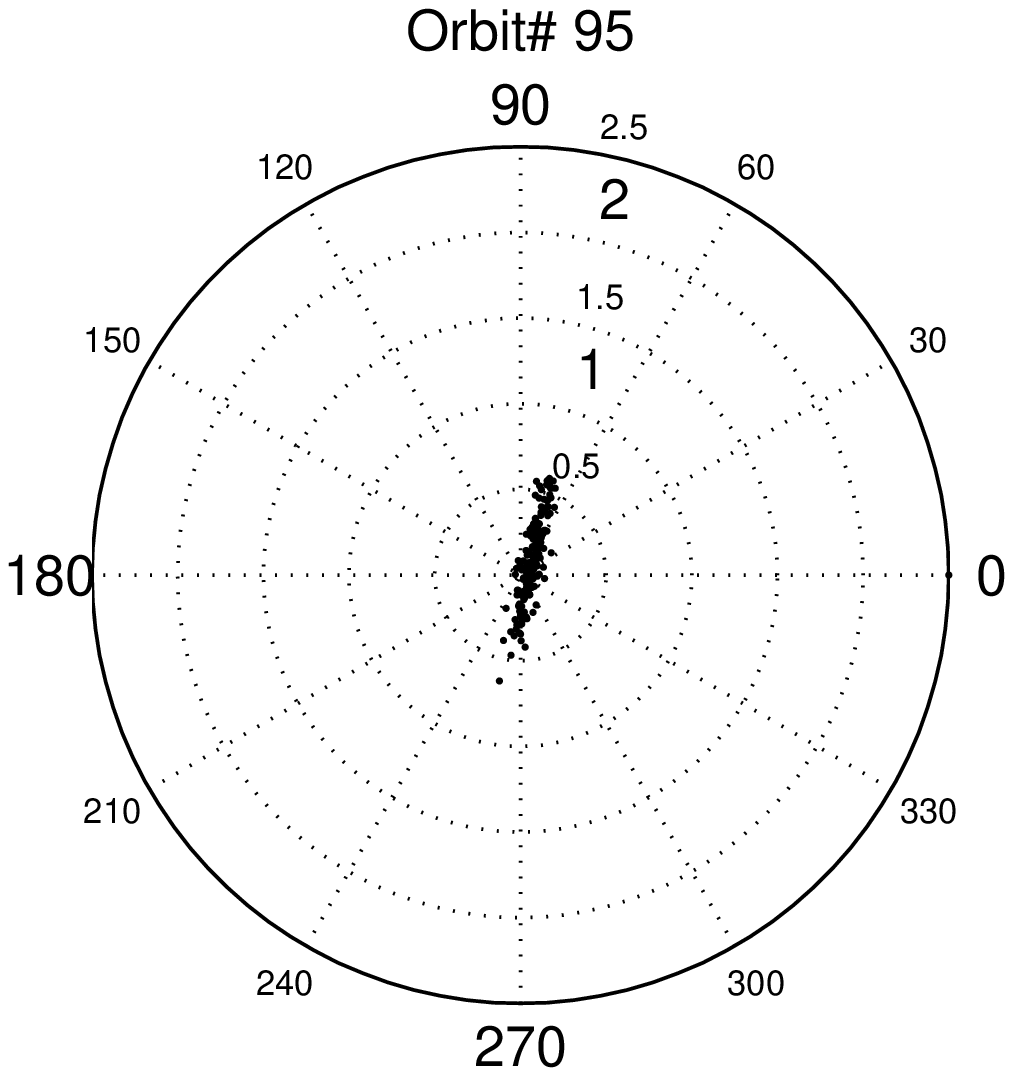}&\includegraphics[scale=0.4, trim = 3mm 0mm 0mm 0mm,
 clip]{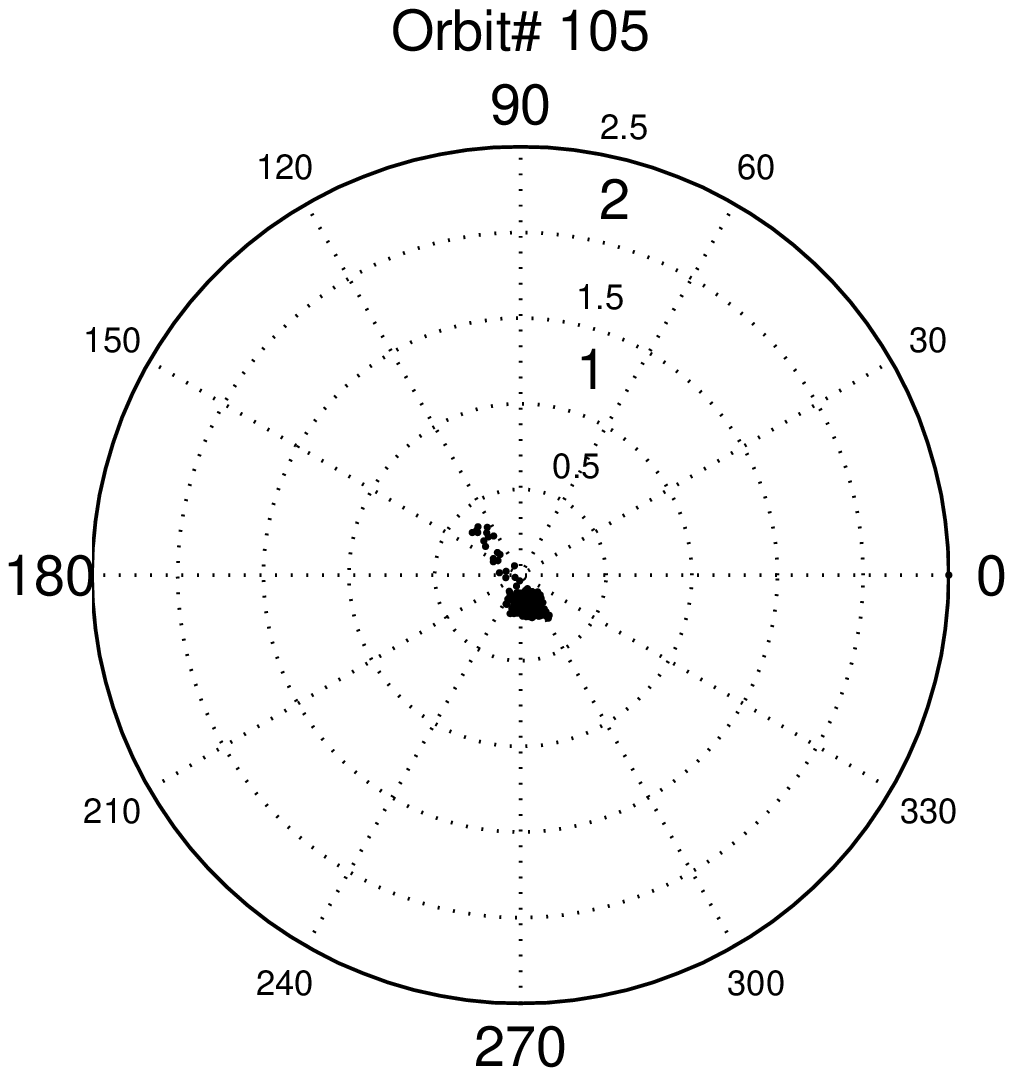}\\
		\end{tabular}
  	\caption{Positions of spin axis in the S/C reference system shown as polar plots for a few example orbits.}
 		\label{figSSAxisPerOrbit}
		\end{figure}

Determinations of IBEX spin axis from all suitable spin blocks in the telemetry obtained from Star Sensor data were performed in the spacecraft polar reference system. A few of them are shown in Fig. \ref{figSSAxisPerOrbit}. The Z-axis coincides with the spin axis pointing determined by the ISOC from ACS data. 		


\end{document}